\documentclass[oldversion]{aa}  
\usepackage{graphicx,amsmath}
\usepackage{color}
\usepackage{natbib}		%for A&A : reimplements the LaTeX \cite command 
\usepackage{url}
\usepackage{multirow}
\bibpunct{(}{)}{;}{a}{}{,} % to follow the A&A style

\begin{document}

   \title{3D model of hydrogen atmospheric escape from HD\,209458b and HD\,189733b: radiative blow-out and stellar wind interactions}
   				   
   \author{
   V.~Bourrier\inst{1,2}\and
   A.~Lecavelier des Etangs\inst{1,2}
	}
   
\authorrunning{V.~Bourrier \& A.~Lecavelier des Etangs}
\titlerunning{Evaporation model}

\offprints{V.B. (\email{bourrier@iap.fr})}

   \institute{
   CNRS, UMR 7095, 
   Institut d'astrophysique de Paris, 
   98$^{\rm bis}$ boulevard Arago, F-75014 Paris, France
   \and
   UPMC Univ. Paris 6, UMR 7095, 
   Institut d'Astrophysique de Paris, 
   98$^{\rm bis}$ boulevard Arago, F-75014 Paris, France
   }
   
   \date{} %Received ...; accepted ...}
 
  \abstract
{
  % context heading (optional)
%   {}
  % aims heading (mandatory)
%   {
Transit observations in the Lyman-$\alpha$ line of the hot-Jupiters HD\,209458b and HD\,189733b revealed strong signatures of neutral hydrogen escaping the planets' upper atmospheres. Here we present a 3D particle model of the dynamics of the escaping atoms. This model is used to calculate theoretical Lyman-$\alpha$ absorption line profiles, which can be directly compared with the absorption observed in the blue wing of the line during the planets' transit.\\
For HD\,209458b, the observed velocities of the planet-escaping atoms up to -130\,km\,s$^{-1}$ are naturally explained by radiation-pressure acceleration. The observations are well-fitted with an ionizing flux of about 3 -- 4 times the solar value and a hydrogen escape rate in the range $10^{9}$ -- $10^{11}$\,g\,s$^{-1}$, in agreement with theoretical predictions.\\
For HD\,189733b, absorption by neutral hydrogen has been observed in September 2011 in the velocity range -230 to -140\,km\,s$^{-1}$.
These velocities are higher than for HD\,209458b and require an additional acceleration mechanism for the escaping hydrogen atoms, which could be interactions with stellar wind protons. We constrain the stellar wind (temperature $\sim3\times10^{4}\,K$, velocity 200$\pm$20\,km\,s$^{-1}$ and density in the range $10^{3}$ -- $10^{7}$\,cm$^{-3}$) as well as the escape rate ($4\times10^{8}$ -- $10^{11}$\,g\,s$^{-1}$) and ionizing flux (6 -- 23 times the solar value). We also reveal the existence of an 'escape-limited' saturation regime in which most of the gas escaping the planet interacts with the stellar protons. In this regime, which occurs at proton densities above $\sim3\times10^{5}$\,cm$^{-3}$, the amplitude of the absorption signature is limited by the escape rate and does not depend on the wind density. The non-detection of escaping hydrogen in earlier observations in April 2010 can be explained by the suppression of the stellar wind at that time, or an escape rate of about an order of magnitude lower than in 2011.\\ 
For both planets, best-fit simulations show that the escaping atmosphere has the shape of a cometary tail. Simulations also revealed that the radiative blow-out of the gas causes spectro-temporal variability of the absorption profile as a function of time during and after the planetary transit. Because no such variations are observed when the absorbing hydrogen atoms are accelerated through interactions with the stellar wind, this may be used to distinguish between the two scenarios.

  % methods heading (mandatory)
%{
%
%}
  % results heading (mandatory)
%{
%   }
  % conclusions heading (optional), leave it empty if necessary 
%   {}
}


   \maketitle

\section{Introduction}
\label{intro} 

Transit observations in the Lyman-$\alpha$ line of neutral hydrogen H\,{\sc i} with the Space Telescope Imaging Spectrograph (STIS) onboard the Hubble Space Telescope (HST) led to the first detection of atmospheric escape from an exoplanet, the hot-Jupiter HD\,209458b (\citealt{VM2003}). The interpretation of these observations has been the subject of debate (\citealt{BJ2007}; \citealt{BJ2008}; \citealt{VM2008}). However, the extended cloud of neutral hydrogen surrounding the planet was confirmed by subsequent observations at low spectral resolutions (\citealt{VM2004}; \citealt{Ehrenreich2008}; \citealt{BJ_Hosseini2010}). Heavier elements such as O\,{\sc i}, C\,{\sc ii}, or Si\,{\sc iii} have also been detected at high altitude from HD\,209458b (\citealt{VM2004}; \citealt{BJ_Hosseini2010}; \citealt{Linsky2010}), suggesting the atmosphere is in a 'blow-off' state. Closer to the Earth than HD\,209458b and orbiting a brighter star, the very hot-Jupiter HD\,189733b came as a particularly good candidate to study atmospheric evaporation (\citealt{Bouchy2005}). Transits have been observed in the Lyman-$\alpha$ line with the Advanced Camera for Surveys (HST/ACS) in 2007-2008 (\citealt{Lecav2010}) and with HST/STIS in September 2011 ({\it e.g.}, \citealt{Bourrier2013}). This revealed the evaporation of neutral hydrogen from the planet's upper atmosphere. Together with the non-detection of atmospheric escape in April 2010, these observations led to the conclusion that there are significant temporal variations in the evaporating atmosphere of HD\,189733b (\citealt{Lecav2012}). Recently, \citet{BJ_Ballester2013} also detected neutral oxygen, and possibly ionized carbon, in the extended atmosphere of HD\,189733b.\\
On the theoretical side, a broad variety of models has been developed to characterize the structure of the upper atmosphere of such close-in giant exoplanets and explain the evaporation process (\citealt{Lammer2003}; \citealt{Lecav2004}, \citeyear{Lecav2007}, \citeyear{Lecav2008b}; \citealt{Baraffe2004,Baraffe2005,Baraffe2006}; \citealt{Yelle2004,Yelle2006}; \citealt{Jaritz2005}; \citealt{Tian2005}; \citealt{GarciaMunoz2007}; \citealt{Holmstrom2008}; \citealt{Stone2009}; \citealt{MurrayClay2009}; \citealt{Adams2011}; \citealt{Guo2011,Guo2013}; \citealt{Koskinen2012a,Koskinen2012b}). Most of these models provide theoretical predictions for the escape rate of hydrogen from hot-Jupiters such as HD\,209458b and HD\,189733b in the range $10^{9}-10^{11}$\,g\,s$^{-1}$. \\
On the observational side, only a few models were developed to directly interpret the observations and estimate the physical conditions in the atmospheres of these two exoplanets by fitting the data. \citet{Schneiter2007} developed a 3D model of the hydrodynamical interaction between the gas escaping from HD\,209458b and the stellar wind, neglecting the photoionization of the neutral hydrogen. Alternatively, in the 3D model developed by \citet{Holmstrom2008} and \citet{Ekenback2010} the production of energetic neutral atoms through the interaction of the planetary and stellar winds at high altitude from the planet is presented as the source for the observed hydrogen. In these 3D models the escaping gas is shaped into a cometary cloud. \citet{Tremblin2012} further explored the process of charge exchange in a 2D hydrodynamic model, focusing more specifically on the mixing layer at the meeting of the two winds. All these models either neglected or underestimated the effect of the stellar radiation pressure on the dynamics of the escaping hydrogen.\\

Here we present a 3D numerical particle model developed to simulate the escape of gas from the atmosphere of an exoplanet. It takes into account the roles of radiation pressure, ionization, self-shielding, and stellar wind interactions. Earlier versions of this model were used by \citet{VM_lecav2004} for HD\,209458b, and by \citet{Lecav2010} for HD\,189733b. In Section~\ref{evap model} we explain how the model processes the dynamics of the escaping atoms and calculates the resulting theoretical absorption profile. In Section~\ref{model results} we apply the model to the hot-Jupiters HD\,209458b and HD\,189733b, and compare simulated spectra with the absorption signatures detected in the blue wing of the resolved Lyman-$\alpha$ line observed with STIS. We constrain the planetary escape rate, the ionizing flux from the star and the stellar wind properties, and describe their influence on the structure of the extended exosphere, and on its temporal variations for HD\,189733b. We discuss the short timescale spectro-temporal variability of the theoretical absorption profile in Section~\ref{spec_temp} and sum up our work in Section~\ref{conclu}.

%%%%%%%%%%%%%%%%%%%%%%%%%%%%%%%%%%%%%%%%%%%%%%%%%%%%%%%%%%%%%%%%%%%%%%%

\section{Evaporation model}
\label{evap model}

\subsection{Overview}

We used Monte-Carlo particle simulations to compute the dynamics of the escaping gas. We considered neutral hydrogen only, even if the model can be used for heavier atoms as well (Vidal-Madjar et al. 2013 in prep.). The hydrogen atoms are represented by meta-particles, and the number of atoms in a meta-particle is calculated to keep a low value of the corresponding optical depth $d\tau$ in front of a fraction of the occulted stellar surface. The total number of particles $dN$ launched every time step \textit{dt} depends on the \textit{escape rate of neutral hydrogen} $\dot{M}$ (first free parameter of the model), so that 
\begin{equation} 
\label{eq:deltaN}
dN=\frac{\dot{M}\,dt}{N_{\mathrm{meta}}\,m_{\mathrm{H}}},
\end{equation}
with $N_{\mathrm{meta}}$ the number of hydrogen atoms in a meta-particle and $m_{\mathrm{H}}$ the mass of a hydrogen atom.\\
The particles are released from the entire upper atmosphere of the planet at a fixed altitude $R_{\mathrm{launch}}$ in between the planetary radius and the Roche lobe. In doing so, we assumed that the strong X/EUV heating on the day-side of the tidally locked hot-Jupiters is redistributed throughout the entire upper atmosphere. The initial velocity distribution of the particles (with respect to the planet) is isotropic, with a thermal speed $v_{\mathrm{therm}}$ corresponding to a 11\,000\,K exobase for HD\,209458b and 14\,000\,K for HD\,189733b ({\it e.g.},  \citealt{Lecav2004}; \citealt{BallesterSing2007}; \citealt{GarciaMunoz2007}; \citealt{MurrayClay2009}, \citealt{Stone2009}). In any case, because the final velocities of the atoms are much higher than their initial velocities (Sect.~\ref{hyd_dyn}), the results presented here do not depend much on the assumptions made on the initial conditions on $R_{\mathrm{launch}}$ and $v_{\mathrm{therm}}$.

The dynamics of the gas is described by the spatial and velocity distributions of the particles with time, calculated with a fourth-order Runge-Kutta algorithm in a Cartesian referential centered on the star. The orthogonal referential was defined with the x-axis pointing toward the Earth, the y-axis in the plane of the planetary orbit in the direction of the planet motion at transit time, and the z-axis completing the right-handed referential. All velocities herein are given with respect to the star. The dynamics of the particles in the stellar referential results from the stellar and planetary gravities (\vec{F_{\mathrm{st-grav}}} and \vec{F_{\mathrm{pl-grav}}}), the stellar radiation pressure $-\beta\,\vec{F_{\mathrm{st-grav}}}$ (see \citealt{Lagrange1998} and Sect.~\ref{rad_press}), and the inertial force \vec{F_{\mathrm{in}}} linked to the non-Galilean referential. The total force on a particle is thus

\begin{equation} 
\label{eq:pfd}
\vec{F}=(1-\beta)\,\vec{F_{\mathrm{st-grav}}}\,+\,\vec{F_{\mathrm{pl-grav}}}\,+\,\vec{F_{\mathrm{in}}}.
\end{equation}
The particles are quickly accelerated away from the star by the strong radiation pressure until it is balanced by the stellar gravity, and the particles reach their maximum radial velocity (Sect.~\ref{hyd_dyn}). This radiation-pressure limited velocity may be exceeded because of additional acceleration mechanisms such as the interaction of the neutral hydrogen gas with stellar wind protons (Sect.~\ref{ENA}). The hydrogen atoms lifetime is limited by the \textit{ionizing stellar extreme-UV (EUV) flux} (second free parameter of the model). The probability that a meta-particle is ionized during a time step \textit{dt} is given by
\begin{align}
\label{eq:ion}
dP&=1-\exp(- \frac{dt}{t_{\mathrm{ion}}}) \\
t_{\mathrm{ion}}&=t_{\mathrm{Sun}}\,\frac{a_{\mathrm{pl}}^2}{F_{\mathrm{ion}}}, \nonumber
\end{align}
with $t_{\mathrm{ion}}$ the characteristic lifetime of a hydrogen atom subject to photoionization, $a_{\mathrm{pl}}$ the semi-major axis of the planet orbit, $F_{\mathrm{ion}}$ the ionizing flux, and $t_{\mathrm{Sun}}\approx150\,days$ the lifetime at 1\,AU assuming an ionization flux corresponding to a solar minimum (\citealt{Bzowski2008}). \\
Hydrogen atoms are removed from the simulation when they become ionized, because they are not visible in Lyman-$\alpha$ anymore. We did not take into account collisions between neutral atoms in the hydrogen cloud. Indeed, the highest densities in our simulations reach $\sim10^{13}$\,m$^{-3}$ at the launching radius, and with a $H-H$ cross section $\sigma\sim10^{-21}\,m^{2}$, the mean free path of a hydrogen atom is about $10^{5}$\,km.

In addition to the neutral hydrogen escape rate ($\dot{M}$, in g\,s$^{-1}$) and the stellar ionizing flux ($F_{\mathrm{ion}}$, in units of solar flux $F_\odot$), three more free parameters are needed when interactions with stellar wind protons are taken into account. These parameters are the \textit{bulk velocity} ($V_{\mathrm{wind}}$, in km\,s$^{-1}$), \textit{temperature} ($T_{\mathrm{wind}}$, in K) and \textit{density} ($n_{\mathrm{wind}}$, in cm$^{-3}$) of the proton distribution at the orbit of the planet. The physical parameters used in the simulations and the numerical parameters with constant values are given in Table~\ref{num_param}.

\begin{table}[tbh]										
\begin{tabular}{llcccc}
\hline
\hline
\noalign{\smallskip}
Parameters 					  & Symbol  													   & \multicolumn{2}{c}{Value}        																							\\
				   						&								  									   & HD\,209458b														& HD\,189733b  													\\
\noalign{\smallskip}
\hline
\noalign{\smallskip}
Distance from Earth		& $D_{\mathrm{*}}$								 	 	 &    47.0\,pc        										&	19.3\,pc										\\
\noalign{\smallskip}
Star radius						& $R_{\mathrm{*}}$								 		 &    1.146$\,R_{\mathrm{Sun}}$        	  &	0.755$\,R_{\mathrm{Sun}}$										\\
\noalign{\smallskip}
Star mass							& $M_{\mathrm{*}}$								 		 &    1.148$\,M_{\mathrm{Sun}}$        		&	0.820$\,M_{\mathrm{Sun}}$										\\
\noalign{\smallskip}
Planet radius					& $R_{\mathrm{pl}}$										 &    1.380$\,R_{\mathrm{Jup}}$       		&	1.138$\,R_{\mathrm{Jup}}$ 									\\
\noalign{\smallskip}
Planet mass						& $M_{\mathrm{pl}}$								 		 &    0.69$\,M_{\mathrm{Sun}}$       			&	1.15$\,M_{\mathrm{Sun}}$										\\
\noalign{\smallskip}
Orbital period				& $T_{\mathrm{pl}}$										 &    3.52475$\,days$        				    	&	2.21858$\,days$ 										\\
\noalign{\smallskip}
Semi-major axis				& $a_{\mathrm{pl}}$								 		 &    0.047$\,AU$        									&	0.031$\,AU$										\\
\noalign{\smallskip}
Inclination					  & $i_{\mathrm{pl}}$								 		 &    $86.59^{\circ}$        							&	$85.51^{\circ}$										\\
\noalign{\smallskip}
\hline
\noalign{\smallskip}
Time step							&  $dt$																 &  \multicolumn{2}{c}{$\approx$300\,s}   																				\\
\noalign{\smallskip}
Velocity resolution 	& $\Delta v$     											 &  \multicolumn{2}{c}{20\,km\,s$^{-1}$}  																				\\
\noalign{\smallskip}
Elementary 						& \multirow{2}{*}{$d\tau$}   					 &  \multicolumn{2}{c}{\multirow{2}{*}{0.05}}     																\\
Optical depth					&  				                   					 &	 																																							\\
\noalign{\smallskip}
Number of atoms       & \multirow{2}{*}{$N_{\mathrm{meta}}$} & \multirow{2}{*}{$4.75\times10^{31}$}   &  \multirow{2}{*}{$2.06\times10^{31}$} \\
Per meta-particle     &																			 &																				&	  																		\\		
\noalign{\smallskip}
Thermal speed					& $v_{\mathrm{therm}}$								 &    13.5\,km\,s$^{-1}$        					&	15.3\,km\,s$^{-1}$										\\
\noalign{\smallskip}
Launching radius			& $R_{\mathrm{launch}}$								 &    2.80$\,R_{\mathrm{pl}}$  						&   2.95$\,R_{\mathrm{pl}}$   					\\
\noalign{\smallskip}
\hline
\hline
\end{tabular}
\caption{Physical parameters for HD\,209458b and HD\,189733b, and numerical parameters with fixed values for all simulations.}
\label{num_param}
\end{table}

%%%%%%%%%%%%%%%%%%%%%%%%%%%%%%%%%%%%%%%%%%%%%%%%%%%%%%%%%%%%%%%%%%%%%%%

\subsection{Acceleration mechanisms}
\label{acc_mecha}

\subsubsection{Radiation pressure}
\label{rad_press}

\begin{figure}[tbh]
\includegraphics[width=\columnwidth]{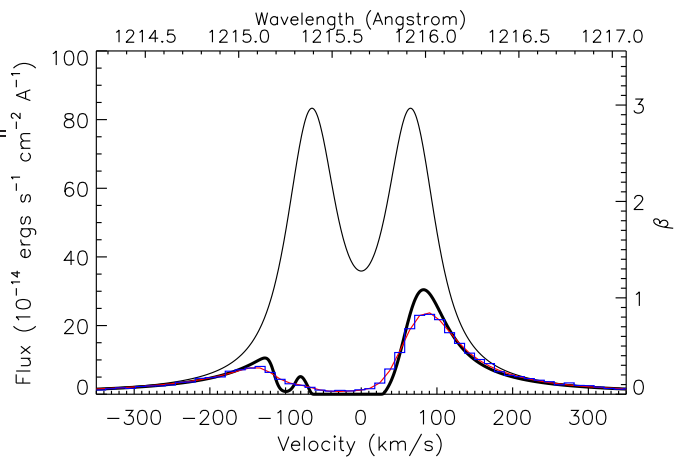}
\caption[]{Theoretical profile of HD\,189733 Lyman-$\alpha$ line. The black thin line shows the theoretical intrinsic stellar emission line profile as seen by hydrogen atoms escaping the planetary atmosphere, scaled to the Earth distance. It also corresponds to the profile of the ratio $\beta$ between radiation pressure and stellar gravity (values reported on the right axis). The black thick line shows the Lyman-$\alpha$ line profile after absorption by the interstellar hydrogen (1215.6\AA) and deuterium (1215.25\AA). The line profile convolved with the STIS G140M instrumental line spread function (red line) is compared with the observations (blue histogram), yielding a good fit with a $\chi^2$ of 36.1 for 40 degrees of freedom. Note that $\beta$ is higher than 1 for velocities in the range -120 to 120\,km\,s$^{-1}$.}
\label{fit_HD189}
\end{figure}

\begin{figure}[tbh]
\includegraphics[width=\columnwidth]{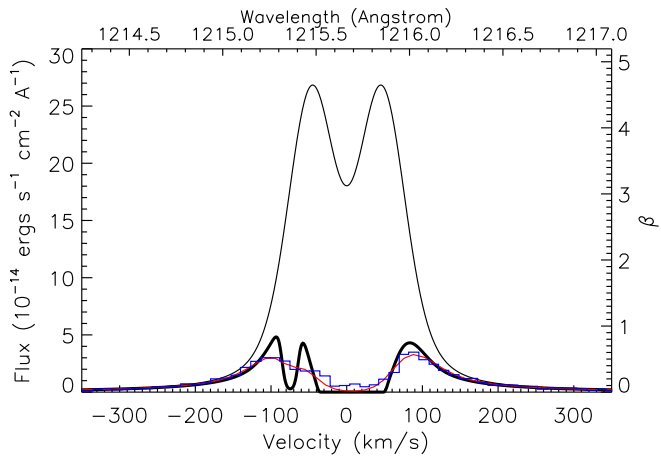}
\caption[]{Same as in Fig~\ref{fit_HD189} for HD\,209458. The ISM deuterium absorption occurs at about 1215.35\AA. The fit with the observations yields a $\chi^2$ of about 49.36 for 31 degrees of freedom. $\beta$ is higher than 1 for velocities in the range -110 to 110\,km\,s$^{-1}$.}
\label{fit_HD209}
\end{figure}

To calculate the radiation pressure applied to each particle, we estimated the coefficient $\beta$ (see Equation~\ref{eq:pfd}) as a function of velocity. This coefficient is indeed proportional to the stellar Lyman-$\alpha$ line flux received by an escaping atom at its radial velocity (with respect to the star), and we thus need to know the Lyman-$\alpha$ line profile as seen by the particle, i.e., without absorption by the interstellar medium. \\
For both HD\,209458 and HD\,189733, we calculated an analytical profile for the stellar Lyman-$\alpha$ line that we compared with the profile observed with the HST/STIS spectrograph (see Sect.~\ref{observations} for more information). The resulting line profile is composed of two peaks separated by a deep absorption due to the interstellar neutral hydrogen H\,{\sc i} in a narrow band at the line center ($\lambda_{\mathrm{0}}=$1215.6\,\AA) and a shallower absorption by the interstellar deuterium D\,{\sc i} blueward of the line center. The Lyman-$\alpha$ line profile can be reconstructed by fitting these observations following the method of \citet{Wood2005}, \citet{Ehrenreich2011} and \citet{France2012}. For HD\,189733 we used the profile estimated by \citet{Bourrier2013}. These authors found that the best fit to the data was obtained with two similar Voigt profiles, as demonstrated using two different information criteria, the Bayesian information criterion (BIC) and the Akaike information criterion (AIC) (see, e.g., de Wit et al. 2012). The comparison between their best model-reconstructed Lyman-$\alpha$ stellar profile and the observed spectrum yields a $\chi^2$ of 36.1 for 40 degrees of freedom in the wavelength range 1214.25-1215.5\,\AA\ and 1215.8-1217.1\,\AA\ (Fig.~\ref{fit_HD189}). We made the same analysis for HD\,209458 and found that the best result was obtained by modeling the stellar emission line with the same profile as for HD\,189733, albeit with different parameter values. The resulting $\chi^2$ yields a value of 49.36 for 28 degrees of freedom in the wavelength range 1214.4-1215.5\,\AA\ and 1215.8-1216.8\,\AA\ (Fig.~\ref{fit_HD209}). The corresponding profile is remarkably similar to the profile found by \citet{Wood2005}.\\
With the reconstructed stellar Lyman-$\alpha$ line profile, we can calculate the radiation pressure on hydrogen atoms as a function of their radial velocity. We found that radiation pressure is higher than the stellar gravity (i.e. $\beta>1$) for all atoms with absolute radial velocities below 120\,km\,s$^{-1}$ for HD\,189733, and 110\,km\,s$^{-1}$ for HD\,189733 (Figs.~\ref{fit_HD189} and \ref{fit_HD209}). Hydrogen atoms escaping the planets' atmospheres are therefore accelerated away from the star to velocities above at least 100\,km\,s$^{-1}$, simply because of radiation pressure. The influence of this force on the hydrogen atoms dynamics is discussed in more detail in Sect.~\ref{hyd_dyn}.

%%%%%%%%%%%%%%%%%%%%%%%%%%%%%%%%%%%%%%%%%%%%%%%%%%%%%%%%%%%%%%%%%%%%%%%

\subsubsection{Acceleration by stellar wind protons}
\label{ENA}

For HD\,189733b, the most recent absorption signature detected in the blue wing of the Lyman-$\alpha$ line  is due to hydrogen atoms moving away from the star at velocities higher than the maximum velocities reached with radiation pressure only (see \citealt{Lecav2012}, \citealt{Bourrier2013} and Sect.~\ref{hyd_dyn}). Stellar wind interactions may be the source for an additional acceleration of the escaping hydrogen atoms at large distances from the planet (\citealt{Holmstrom2008}; \citealt{Ekenback2010}; \citealt{Tremblin2012}). A stellar wind proton may gain an electron from the interaction with a neutral hydrogen atom escaping the planet's upper atmosphere. The former proton thus becomes a neutral hydrogen atom (the so-called energetic neutral atoms). We handled the process of charge-exchange as an impulse given to the escaping neutral hydrogen atom, so that it takes the velocity of the incoming proton. With the assumption that there is neither deflection nor energy loss, the population of accelerated hydrogen atoms contributes to the absorption in the Lyman-$\alpha$ line at the high velocities of the proton distribution of the stellar wind . We characterized the stellar wind at the orbit of the planet by its Maxwellian speed distribution (defined by the radial bulk velocity and thermal width) and its proton density.\\
The probability that a neutral hydrogen atom is accelerated by a proton during the time \textit{dt} is given by

\begin{equation}
dP = 1-\exp(- \sigma_{\mathrm{H - H^{+}}} \, ||  \vec{V_{\mathrm{H}}} - \vec{V_{\mathrm{H^{+}}}} \| \, n_{\mathrm{H^{+}}}  \,  dt),
\label{eq:ENA}
\end{equation}
with $n_{\mathrm{H^{+}}}$ the proton density in the vicinity of the hydrogen atom, \vec{V_{\mathrm{H}}} and \vec{V_{\mathrm{H^{+}}}} the respective velocities of the hydrogen atom and the interacting proton, and $\sigma_{H - H^{+}}$ the cross-section of their interaction. For energies in the order of 200\,eV (associated to a bulk velocity $V_{wind}$ in the order of 200\,km\,s$^{-1}$, see Sect.~\ref{HD189}) the energy-dependent cross-section is about $2\times10^{-15}$\,cm$^{2}$ (Fig.~1 of \citealt{Lindsay2005})\\

%%%%%%%%%%%%%%%%%%%%%%%%%%%%%%%%%%%%%%%%%%%%%%%%%%%%%%%%%%%%%%%%%%%%%%%

\subsection{Self-shielding}
\label{self-shielding}

\subsubsection{Stellar photons}

Our model takes into account the self-shielding of the neutral hydrogen cloud, which leads to the extinction of the photons responsible for the radiation pressure and the ionization when they cross the cloud. The photons emitted by a stellar surface element propagate inside the 'cone' centered on the surface normal. As they penetrate deeper into the cloud, the number of photons inside the cone decreases. As a result, fewer hydrogen atoms are accelerated by radiation pressure in the layers farther away from the star, since they are shielded by the gas layers closer to the star. Inside the cone, the coefficient $\beta$ at the velocity \textit{v} decreases as a function of the penetration depth \textit{$\Lambda$} of the photons into the cloud, 

\begin{align}
\label{eq:eqbeta}
&\beta(v,\Lambda)=\beta(v,0) \, \exp(- \frac{\sigma_{\mathrm{v0}} \, \lambda_{0}}{\Delta v} \, \int\limits_{\mathrm{0}}^{\mathrm{\Lambda}}{n_{\mathrm{H}}(\mu,v) \, d\mu}) \\
&\mathrm{where}~\sigma_{\mathrm{v0}}=\frac{\pi \, e^{2}}{4 \, \pi \, \epsilon_{0} \, m_e \, c}\,f_{osc}, \nonumber
\end{align}

with $n_{\mathrm{H}}$ the density of hydrogen gas inside the cone at a given depth and taking into account all atoms moving at the velocity $v\pm\Delta v/2$. $\sigma_{\mathrm{v0}}$ is the normalization constant of the absorption cross-section profile for neutral hydrogen (in units of $m^2$\,s$^{-1}$, see Table~\ref{dampingval}), and $f_{osc}$ the line oscillator strength.

Similarly, the model takes into account the self-extinction of the photons responsible for the ionization of the hydrogen atoms. However, whereas self-shielding is critically important for radiation pressure (Sect.~\ref{self_views}), it has a negligible impact on the lifetimes of the neutral hydrogen atoms, because of the smaller cross-section for their photoionization, $\overline{\sigma}_{ion}$, calculated as 

\begin{equation}
\label{eq:eqphotoion}
\overline{\sigma}_{ion}=\frac{\int_{\lambda}{\sigma_{ion}(\lambda)\,f_{ion}(\lambda)\,d\lambda}}{\int_{\lambda}{f_{ion}(\lambda)\,d\lambda}}, \\
\end{equation}
with $\sigma_{ion}(\lambda)$ and $f_{ion}(\lambda)$ the ionization cross-sections and solar photon flux density at the wavelength $\lambda$, in the range 110 to 912\,\AA. We found $\overline{\sigma}_{ion}\sim5\times10^{-18}\,cm^{2}$ (see \citealt{Hinteregger1960}).\\

\subsubsection{Stellar protons}

Outside of the hydrogen cloud, we assumed that the stellar wind density decreases as a function of the distance from the star according to a quadratic law. Inside the cloud, the hydrogen atoms of the outer regions are shielded from the stellar protons in the same way as they are shielded from the photons. We found that this effect is not negligible. The proton density in the hydrogen cloud decreases with the penetration depth \textit{$\Lambda$} in the same way as $\beta$ in Equation~\ref{eq:eqbeta}, but the optical depth $\tau_\mathrm{H - H^{+}}$ now depends on the relative velocity between hydrogen atoms and protons, 
 
\begin{equation}
\tau_\mathrm{H - H^{+}}(\Lambda) = \frac{\sigma_\mathrm{H - H^{+}}}{V_{wind}} \, \int\limits_{\mathrm{0}}^{\mathrm{\Lambda}}{ n_{\mathrm{H}}(\mu) \, \| \vec{V_{\mathrm{H}}(\mu)} -  \vec{V_{wind}} \| \, d\mu}
\end{equation}

%%%%%%%%%%%%%%%%%%%%%%%%%%%%%%%%%%%%%%%%%%%%%%%%%%%%%%%%%%%%%%%%%%%%%%%

\subsection{Absorption}
\label{abs_model}

The model calculates at each time step the theoretical spectral absorption in the Lyman-$\alpha$ line as observed from the Earth. During the transit, the stellar flux is absorbed at every wavelength of the line by the planetary disk. In addition to this planetary occultation depth, the neutral hydrogen atoms contribute to the absorption at specific wavelengths corresponding to their velocities projected on the star/Earth line of sight. After the end of the planet transit (i.e., after the fourth contact), there may still be a significant absorption that is exclusively due to the hydrogen cloud trailing behind the planet like a cometary tail (Sect.~\ref{com_tail}).

\subsubsection{Planetary occultation and hydrogen optical depth}
We discretized the stellar and planetary disks with square grids by $21\times21$ cells and $201\times201$ cells, respectively. During the transit, we estimated the absorption by the planet body $A_\mathrm{pl}(C)$ for a given stellar cell \textit{C} by calculating the value of the stellar surface that was occulted by the planetary disk.\\
Then we calculated the optical depth of the hydrogen gas along the star/Earth line of sight in front of each stellar cell, leaving out particles in front of, or behind, the planetary disk. The optical depth is calculated as a function of the velocities of the particles in front of the cells, with a resolution $\Delta v$ set to the resolution of STIS observations (about 20\,km\,s$^{-1}$). Because of the natural Lorentzian broadening of the absorption lines, a hydrogen atom absorbs the Lyman-$\alpha$ line flux in an extended wavelength range and not just at the single wavelength corresponding to its velocity. Consequently, the optical depth estimated for a given velocity takes into account all the atoms in front of a stellar cell, whatever their velocities. Note that in this section all velocities are projections on the star/Earth line of sight. The general equation for the optical depth $\tau(C,v)$ at a velocity $v$ in front of cell C is 

\begin{equation}
\label{eq:damping0}
\tau(C,v)= \sigma_{\mathrm{v0}} \int\limits_{\mathrm{-\infty}}^{\mathrm{\infty}}{N_\mathrm{spec}(C,v^{,}) \, \phi(v^{,}-v) \, dv^{,}},			
\end{equation} 
with $N_\mathrm{spec}(C,v)$ the number of neutral hydrogen atoms moving at the velocity \textit{v} in front of cell C per unit of surface and velocity, and $\phi(v)$ the line profile of the absorption (centered on the null velocity) at the velocity $v$.\\

%NB : attention l'integrale de phi(v)dv n'est pas la meme que celle de phi(nu)dnu
%NB: La profondeur optique discretise est tau(V)=Sum(N(V')*sigma(V'))=sum(N(V')*sigma_nu0*Phi_V(V'))
%Le profil spectral est le meme quelque soit la valeur centrale (ici V). On somme la contribution de tous les atomes, ceux a la vitesse V' absorbant la raie a la vitesse V'
%On peut ecrire cette formule sous la forme tau(V)=sigma_nu0*int(N_spec(V')*Phi_V(V')*dV') avec N_spec la densite de colonne spectrale (cad par unite de vitesse)
%Si la raie d'absorption est centree sur V, une particule a la vitesse V' absorbe relativement a la raie a la vitesse V'-V. Plutot que d'ecrire Phi_V(V'), puisque la raie est la meme partout, on peut ecrire Phi(V'-V)

To calculate the absorption spectrum with a resolution $\Delta v$, we discretized the optical depth and estimated its value $\tau_{H}(C,v_{\mathrm{i}})$ at the velocity \textit{v}$_{\mathrm{i}}$ in the interval \textit{v}$_{\mathrm{i}}$$\pm\Delta v/2$, in front of cell \textit{C}. First, it takes into account the mean optical depth due to atoms moving at the velocity \textit{v}$_{\mathrm{i}}$ in front of this cell

\begin{equation}
\label{eq:damping1}
\tau_{\mathrm{line}}(C,v_{\mathrm{i}})=\frac{ \sigma_{\mathrm{v0}} \, N(C,v_{\mathrm{i}}) \, \lambda_{0} } { \Delta v },
\end{equation} 

with $\lambda_{\mathrm{0}}$ the central wavelength of the Lyman-$\alpha$ line.
Then, the contribution to $\tau_{H}(C,v_{\mathrm{i}})$ of all the atoms moving at a velocity $v_{\mathrm{j}}\,\neq\,v_{\mathrm{i}}$ in front of the same cell is approximated by

\begin{equation}
\label{eq:damping2}
\tau_{\mathrm{broad}}(C,v_{\mathrm{i}},v_{\mathrm{j}})=\sigma_{\mathrm{v0}} \, N(C,v_{\mathrm{j}}) \, \frac{\Gamma}{4 \, \pi^2} \, \left(\frac{\lambda_{\mathrm{0}}}{v_{\mathrm{j}}-v_{\mathrm{i}}}\right)^2, 
\end{equation} 
 
with $\Gamma$ the damping constant ({\it e.g.}, the width at mid-height of the Lorentzian broadening profile).
Finally, the total optical depth estimated at the velocity \textit{v}$_{\mathrm{i}}$ in front of cell \textit{C} is 

\begin{equation}
\label{eq:damping3}
\tau_{H}(C,v_{\mathrm{i}})=\tau_{\mathrm{line}}(C,v_{\mathrm{i}})+\sum_{\mathrm{j \neq i}}{\tau_{\mathrm{broad}}(C,v_{\mathrm{i}},v_{\mathrm{j}})}.
\end{equation} 
These calculations are based on the book \textit{Interpreting Astronomical Spectra} by D.Emerson. See Table~\ref{dampingval} for parameters and constant values.

\begin{table}[tbh]
\begin{tabular}{clcccc}
\hline
\hline
\noalign{\smallskip}
Parameters 												& Value  \\
\noalign{\smallskip}
\hline
\noalign{\smallskip}
$\lambda_{\mathrm{0}}$						& 1215.6702\,\AA								\\
\noalign{\smallskip}
$\sigma_{\mathrm{v0}}$ 						& $1.102\times10^{-6}$\,m$^{2}$\,s$^{-1}$ \\
\noalign{\smallskip}
$\Gamma$													& $6.27\times10^{8}$\,s$^{-1}$							\\
\noalign{\smallskip}
\hline
\hline
\end{tabular}
\caption{Values of the parameters and constants used in calculating the optical depth of the neutral hydrogen cloud.}
\label{dampingval}
\end{table}

\subsubsection{Total spectral absorption}
The final quantity we look for is the absorption profile of the Lyman-$\alpha$ flux emitted by the whole stellar disk at a given time. It takes into account the planetary occultation depth (during the transit) and the neutral hydrogen absorption, so that 

\begin{equation}
\label{eq:damping4}
C_{\mathrm{A}}(\lambda)=\frac{1}{N_{\mathrm{cells}}}\sum_{\mathrm{\substack{stellar\\cells}}}(1-A_{\mathrm{pl}}(C)) \, \exp(-\tau_{\mathrm{H}}(C,v_{\lambda})),
\end{equation}
with $C_{\mathrm{A}}(\lambda)$ the absorption coefficient at the wavelength $\lambda$, $N_{\mathrm{cells}}$ the number of cells in front of the stellar disk, $v_{\lambda}$ the Doppler velocity associated to $\lambda$ by the relation \mbox{$\lambda-\lambda_{\mathrm{0}} = v_{\lambda}/c$}, $A_{\mathrm{pl}}(C)$ the planetary absorption in front of the stellar cell \textit{C} and $\tau_{\mathrm{H}}(C,v)$ the optical depth of hydrogen atoms in front of the cell at the velocity $v\pm(\Delta v/2)$. 

Using Equation~\ref{eq:damping4}, we calculated the Lyman-$\alpha$ profile after absorption by the planet and the hydrogen gas. To be compared with real observations, the profile was then attenuated by the ISM absorption and convolved with the instrumental line spread function (see Sect.~\ref{rad_press}).

%%%%%%%%%%%%%%%%%%%%%%%%%%%%%%%%%%%%%%%%%%%%%%%%%%%%%%%%%%%%%%%%%%%%%%%

%%%%%%%%%%%%%%%%%%%%%%%%%%%%%%%%%%%%%%%%%%%%%%%%%%%%%%%%%%%%%%%%%%%%%%%

\section{Modeling the atmospheric escape from HD\,209458b and HD\,189733b}
\label{model results}

\subsection{Lyman-$\alpha$ observations}
\label{observations}

\subsubsection{HD\,209458b}
\label{intro_HD209}

\begin{figure}[tbh]
\includegraphics[angle=-90,width=\columnwidth]{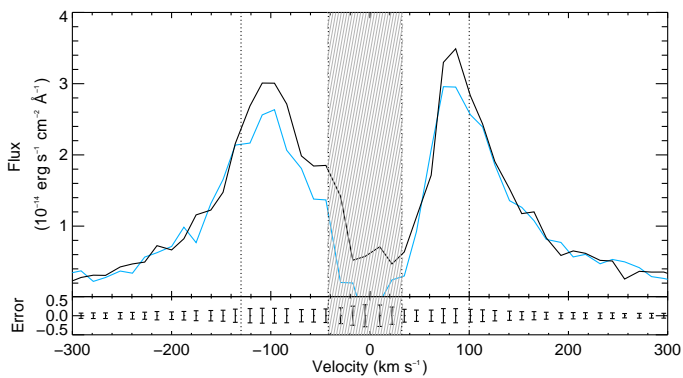}
\caption[]{Lyman-$\alpha$ line as a function of velocity, observed by \citet{VM2003} before (black line) and during (blue line) the transit of HD\,209458b. The striped gray zone corresponds to the region contaminated by the geocoronal emission. The absorption signatures detected in the blue and red wings of the line are delimited by vertical dotted lines. The bottom panel shows the error bars on the flux measured before the transit.} 
\label{absorptionHD209}
\end{figure}

The hot-Jupiter HD\,209458b is located 47 parsecs away from the Earth. It orbits a solar-like star with a semi-major axis of 0.047~AU and an orbital period of 3.5~days. We applied our model to this exoplanet to determine the physical conditions leading to the absorption features detected with the Space Telescope Imaging Spectrograph (STIS) onboard the Hubble Space Telescope (HST) in 2001 with three cumulated transit observations (\citealt{VM2003}). We used these particular data because of their good spectral resolution and the fact that they show the most significant detection of H\,{\sc i} atmospheric escape from this planet (see Sect.~\ref{intro}). The deepest absorption signature was found in the blue wing of the Lyman-$\alpha$ line from -130 to -40\,km\,s$^{-1}$, while a less significant feature was detected in the red wing from 30 to 100\,km\,s$^{-1}$ (Fig.~\ref{absorptionHD209}; see also \citealt{VM2008}). The center of the line does not provide useful information because of the contamination by the geocoronal airglow emission (\citealt{VM2003}). The velocity structure and depth of these absorption features has been much discussed ({\it e.g.}, \citealt{VM2008}, \citealt{BJ2008}). In this paper (Sect.~\ref{HD209}), we fit the spectral absorption depth profile calculated with the spectra given in \citet{VM2008} (Fig.~\ref{absorptionHD209}). The $\chi^2$ of the fits are calculated in the range -300 to -40\,km\,s$^{-1}$ and 30 to 300\,km\,s$^{-1}$, without any assumption on the velocity range for the absorption signatures. In 2001, no observations were carried out after the end of the transit. However, a post-transit observation of the unresolved Lyman-$\alpha$ line, made in 2006 with the Advanced Camera for Surveys (ACS) onboard the HST, showed no absorption with a depth of 1.7$\pm$9\% about 3\,h after the center of the transit (\citealt{Ehrenreich2008}). We discuss in Sect.~\ref{com_tail} how this result may constrain the physical conditions in the atmosphere of HD\,209458b.

\subsubsection{HD\,189733b}
\label{intro_HD189}

\begin{figure}[tbh]
\includegraphics[angle=-90,trim=0cm 0cm 3.91cm 0cm, clip=true,width=\columnwidth]{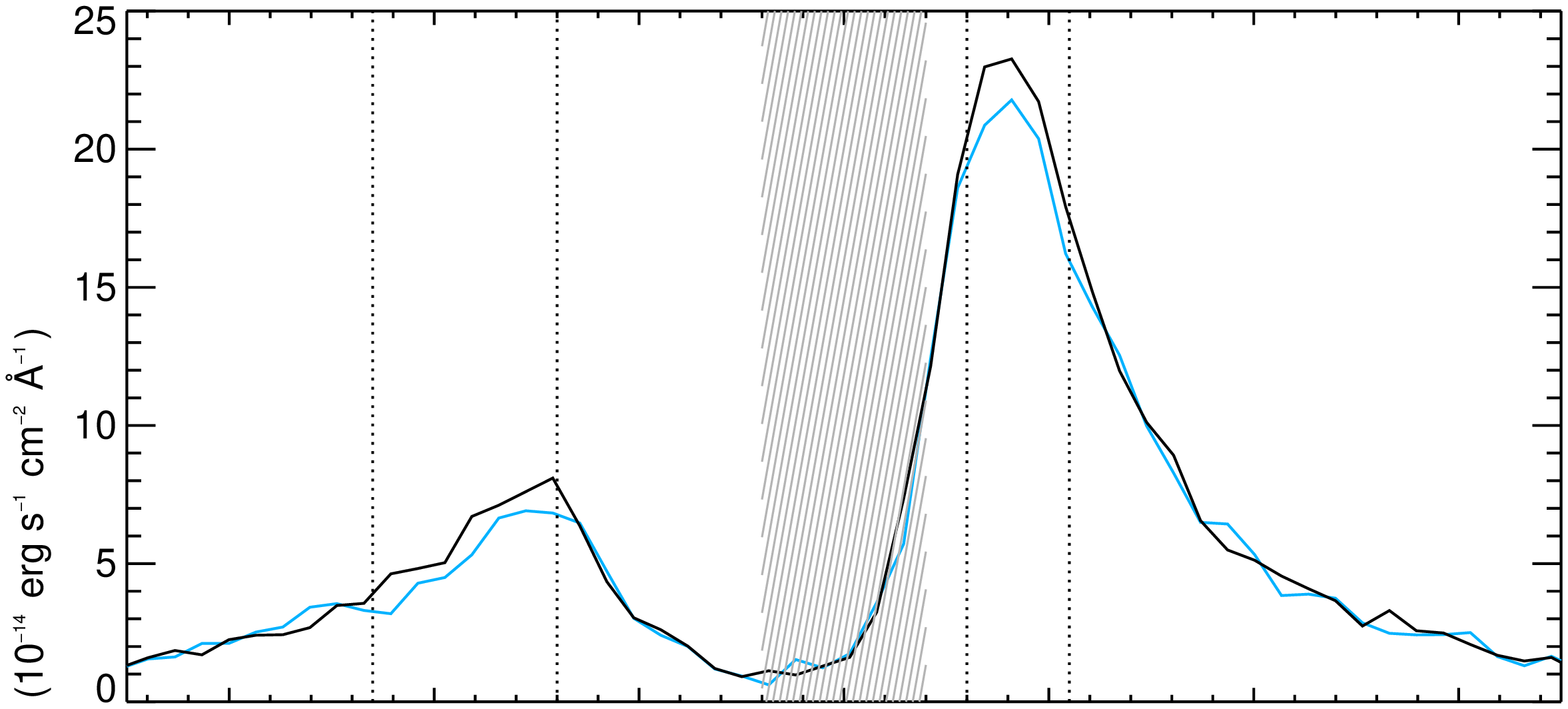}		
\includegraphics[angle=-90,trim=1.3cm 0cm 0cm 0cm, clip=true,width=\columnwidth]{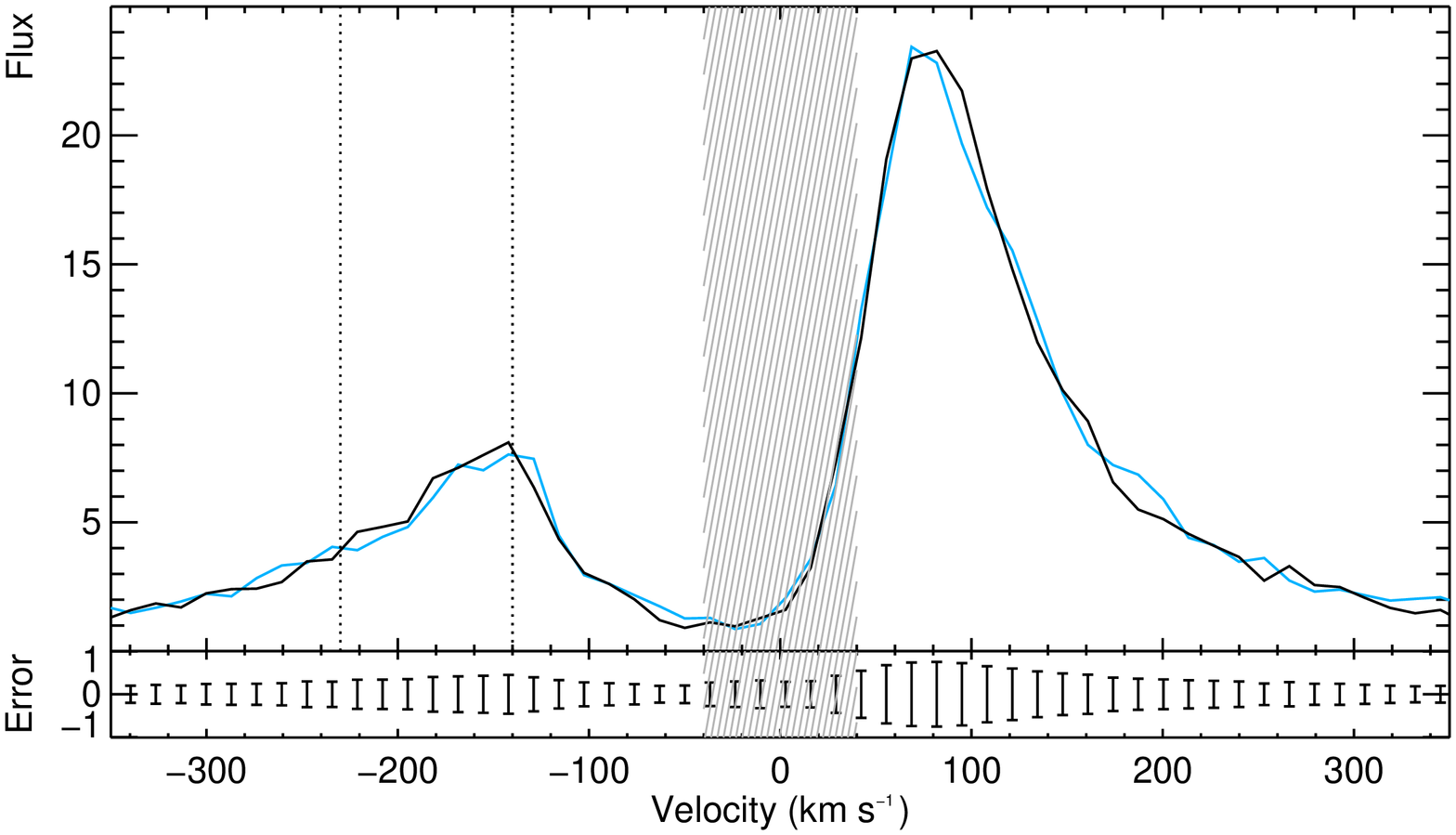}		
\caption[]{Lyman-$\alpha$ line of HD\,189733b as a function of velocity, in September 2011 (\citealt{Lecav2012}; \citealt{Bourrier2013}). The spectrum measured before the planet transit is displayed as a black line, to be compared with the spectra observed during (blue line, top panel) and after (blue line, bottom panel) the transit. The striped gray zone corresponds to the region contaminated by the geocoronal emission. The absorption signatures detected in the blue wing of the line during the transit and the post-transit and in the red wing during the transit are delimited by vertical dotted lines. The bottom panel shows the error bars on the flux measured before the transit.} 
\label{absorptionHD189}
\end{figure}

\begin{figure}[tbh]
\includegraphics[angle=-90,trim=0cm 0cm 3.91cm 0cm, clip=true,width=\columnwidth]{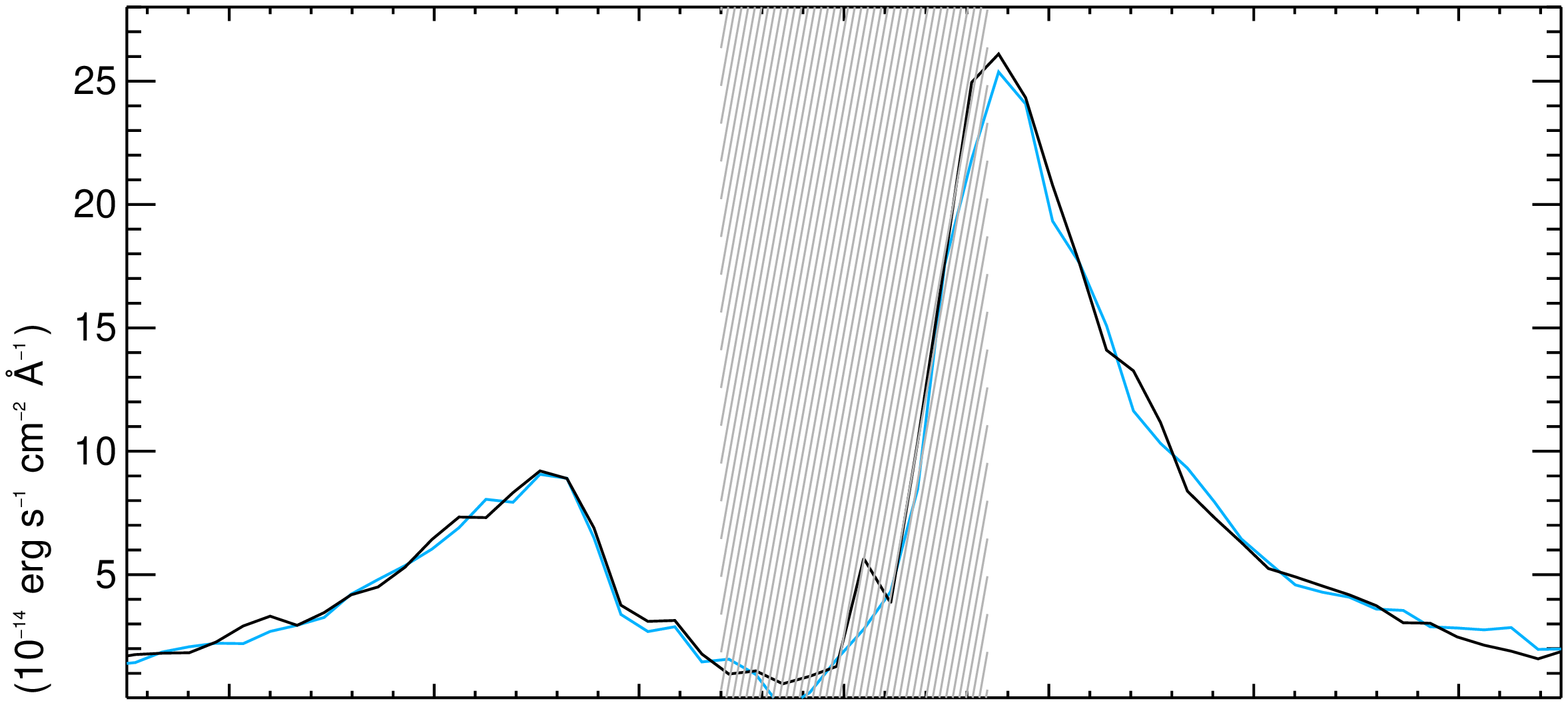}		
\includegraphics[angle=-90,trim=1.3cm 0cm 0cm 0cm, clip=true,width=\columnwidth]{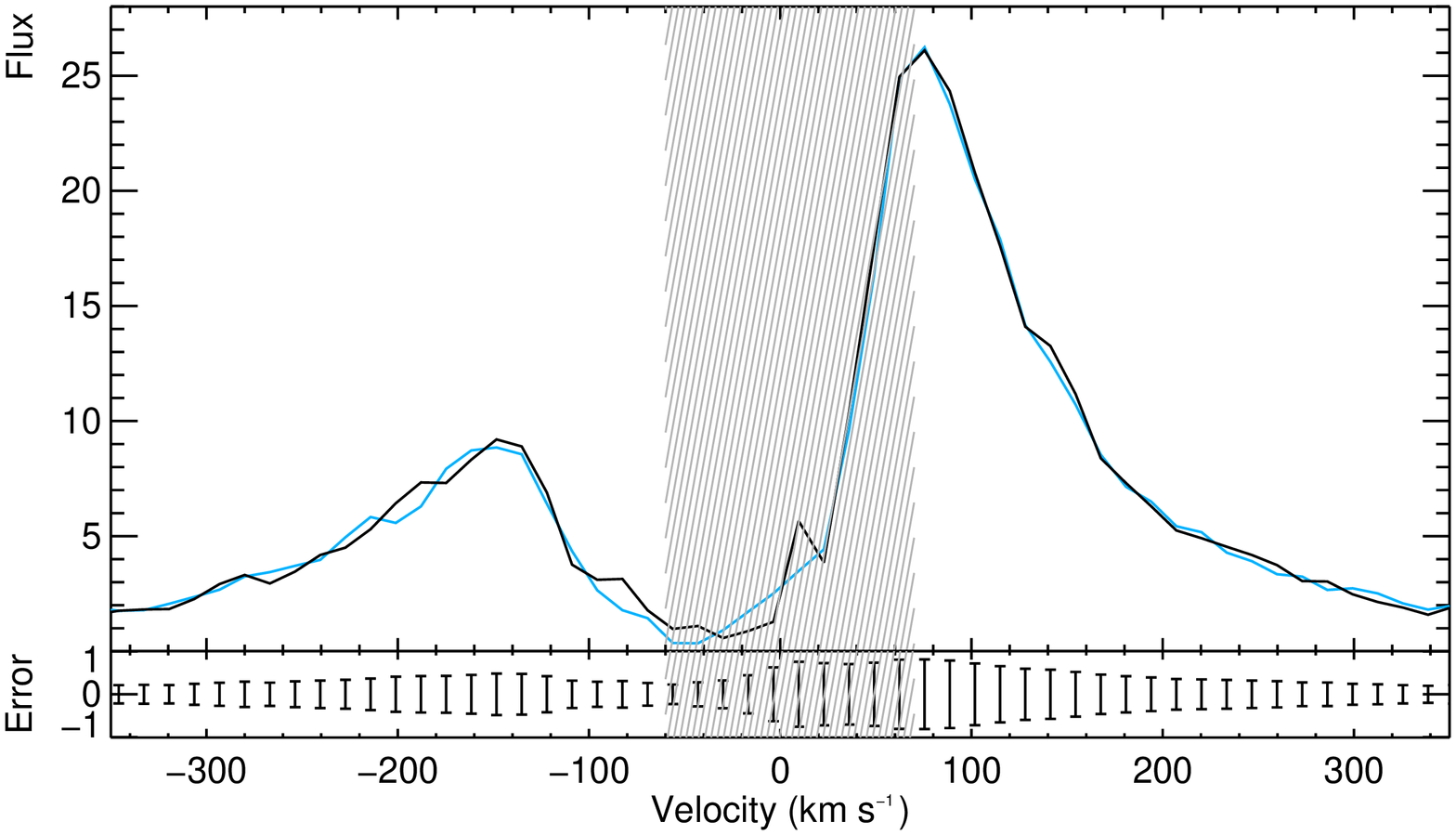}	
\caption[]{Lyman-$\alpha$ line of HD\,189733b as a function of velocity, in April 2010 (\citealt{Lecav2012}). The spectrum measured before the planet transit is displayed as a black line, to be compared with the spectra observed during (blue line, top panel) and after (blue line, bottom panel) the transit. The striped gray zone corresponds to the region contaminated by the geocoronal emission. No absorption signatures were detected in either the transit or post-transit observations. The bottom panel shows the error bars on the flux measured before the transit.}
\label{absorptionHD189_2010}
\end{figure}

We also applied the model to the Lyman-$\alpha$ observations of the very hot-Jupiter HD\,189733b. The planet has a semi-major axis of 0.03~AU and an orbital period of 2.2~days. It is located 19.3~parsecs away from the Earth, and orbits a bright main-sequence star at Lyman-$\alpha$. Temporal variations were detected with HST/STIS in the evaporating atmosphere of HD\,189733b between April 2010 and September 2011 (\citealt{Lecav2012}). To begin with (Sect.~\ref{HD189}), we focused on the physical conditions of the evaporation at the epoch of the second observation, when neutral hydrogen absorption was found in the spectrally resolved Lyman-$\alpha$ line. As for HD\,209458b, two absorption signatures were detected in the wings of the line, the strongest in the range -230 to -140\,km\,s$^{-1}$, whereas a red wing absorption feature was detected from 60 to 110\,km\,s$^{-1}$. Post-transit observations were made during the same 2011 visit with enough sensitivity to find a putative absorption at nearly 2\,$\sigma$ in the range of the signature detected in the blue wing during the transit (\citealt{Bourrier2013}). The $\chi^2$ of the fits to the data were calculated in the range -350 to -50\,km\,s$^{-1}$ and 50 to 350\,km\,s$^{-1}$, taking into account both transit and post-transit observations. All the absorption signatures are located in the wings of the Lyman-$\alpha$ line at higher velocities than for HD\,209458b and are not contaminated by the airglow emission confined in the very center of the line within $\pm40$\,km\,s$^{-1}$ (Fig.~\ref{absorptionHD189}). The high velocities of the atoms producing the absorption in the blue wing cannot be explained by radiation pressure alone, and interactions with stellar wind protons were taken into account in the simulations. In Sect.~\ref{noevap} we estimated constraints on the model parameters for the simulated absorption profile to be consistent with the non-detection of absorption in 2010. Because of the stronger airglow emission at this epoch, the $\chi^2$ were calculated in the range -350 to -60\,km\,s$^{-1}$ and 70 to 350\,km\,s$^{-1}$  (Fig.~\ref{absorptionHD189_2010}). In contrast to \citet{Holmstrom2008} and \citet{Ekenback2010} for HD\,209458b, we did not consider magnetic interactions. Note that the stellar magnetic field was estimated to be about 4 to 23\,mG on average at the orbit of the planet (\citealt{Moutou2007}; \citealt{Fares2010}). \\
\citet{Jensen2012} reported a significant absorption feature at H$\alpha$ in transmission spectra of HD189733b. The comparison between Lyman-$\alpha$ and H$\alpha$ absorption signatures observed at the same epoch would allow for a direct measurement of hydrogen excitation temperature, and a better understanding of the atmospheric escape process. Unfortunately, this comparison cannot be made with the 2011 observations from \citet{Lecav2012}, first because of the evidence for temporal variations in the escaping atmosphere of the planet (the observations from \citealt{Jensen2012} were made between 2006 and 2008), and second because of the different velocity ranges of the absorption features. The hydrogen atoms responsible for the H$\alpha$ absorption were detected within a narrow band at the line center (-24.1 to 26.6\,km\,s$^{-1}$), which is incompatible with the high-velocity population of hydrogen atoms observed beyond -140\,km\,s$^{-1}$ that are responsible for the 2011 Lyman-$\alpha$ absorption.\\

For practical CPU time reasons we limited the exploration of the parameter space by taking the escape rate equal to or below $10^{11}$\,g\,s$^{-1}$. Because of the possibility of detecting spectro-temporal variations at short timescales in the velocity profile of the absorption (see \citealt{Bourrier2013}), we compared the observations of both planets with the spectra calculated for the same time windows as the STIS observations (Table~\ref{obs_log}). For HD\,209458b, because the absorption measured by \citet{VM2003} was obtained by combining the transit observations of three visits, we used a time window that lasts from the start of the earliest exposure (first visit) to the end of the latest exposure (second visit). Since the time-tagged observations of HD\,189733b yielded enough signal for the exposures to be divided into six 379~s subexposures, we chose a time-step of $\approx$300~s for all simulations. Instead of using only the absorption simulated at mid-transit, we averaged the fluxes simulated during the time window of the observations (transit or post-transit) and calculated the resulting mean absorption. In Sect.~\ref{spec_temp} we discuss the variations of the velocity structure of the absorption with time during the atmospheric transit.

\begin{table}[tbh]
\begin{tabular}{p{3cm}cccc}
\hline
\hline
\noalign{\smallskip}
Planet &  \multicolumn{2}{c}{Transit} & \multicolumn{2}{c}{Post-transit}   \\
       &   Start & End 								& Start & End			   \\
\noalign{\smallskip}
\hline
\noalign{\smallskip}
HD\,209458b & -00:35  & 00:54  & \multicolumn{2}{c}{N/A}  \\
\noalign{\smallskip}
HD\,189733b &    &    &  &   \\
\multicolumn{1}{r}{April 2010} & -00:10   &  00:27  & \ 01:25 &  \ 02:03 \\
\multicolumn{1}{r}{September 2011} & -00:26   &  00:11  & \ 01:10 &  \ 01:47 \\
\noalign{\smallskip}
\hline
\hline
\end{tabular}
\caption{Log of the transit and post-transit observations for HD\,209458b in 2001, and HD\,189733b in 2010 and 2011. Time is given in UT and is counted from the center of the transit.}
\label{obs_log}
\end{table}

%%%%%%%%%%%%%%%%%%%%%%%%%%%%%%%%%%%%%%%%%%%%%%%%%%%%%%%%%%%%%%%%%%%%%%%

\subsection{Hydrogen dynamics with radiation pressure}
\label{hyd_dyn}

\subsubsection{Radial velocity observed from the Earth}
\label{vel_from_Earth}

\begin{figure}[tbh]
\centering
\includegraphics[angle=-90,scale=.4]{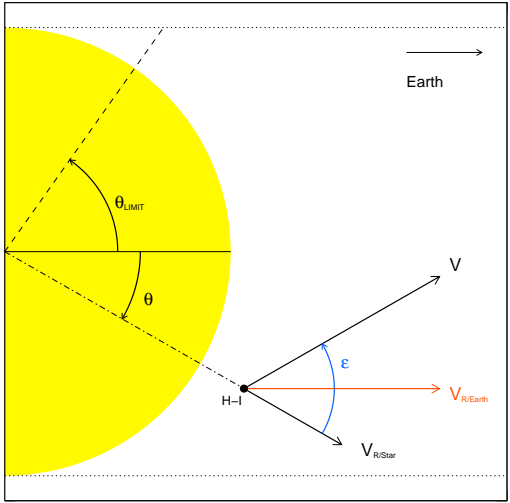}		
\caption[]{Representation of a H\,{\sc i} particle in the orbital plane. $\theta$ is the angular position of the particle, and the horizontal black dotted lines correspond to the maximum values of $\theta$ for which the particle can be observed from the Earth. The radial velocity \vec{V_{\mathrm{r\star}}} and total velocity \vec{V} are displayed as black vectors, separated by the angle $\epsilon$ (positive in our representation). The velocity of the particle as seen from the Earth, \vec{V_{\mathrm{r\oplus}}}, is displayed as a red vector. It is the total velocity vector projected on the star/Earth line of sight. For the sake of clarity, the diagram is not to scale.} 
\label{schema}
\end{figure}

Before fitting the observations, we used a simplified 2D version of the model to make a first analysis of the dynamics of a hydrogen atom escaping one of the planets' upper atmosphere. In this toy-model confined to the planet orbital plane, the particle is launched from the planet with its orbital velocity in the stellar referential. We assumed the planet to be on a circular orbit ($e=4.10^{-4}\pm3.10^{-3}$ for HD\,209458b; $e=4.10^{-3}\pm2.10^{-3}$ for HD\,189733b). The particle trajectory is constrained by the stellar gravity and radiation pressure only. Radiation pressure opposes stellar gravity and accelerates particles away from the star. However, the radiation pressure applied to a particle strongly depends on its radial velocity (in this case the velocity projected on the radial direction with respect to the star). We must point out that even when the particle transits the stellar disk, the radial velocity \textit{relative to the star} ($V_{\mathrm{r\star}}$) may be significantly different from the radial velocity \textit{as observed from the Earth} ($V_{\mathrm{r\oplus}}$). The latter is the velocity projected on the star/Earth line of sight, and corresponds to the wavelengths of the observed spectrum. We can express $V_{\mathrm{r\oplus}}$ as a function of the angular position $\theta$ of the particle in the orbital plane, its radial velocity $V_{\mathrm{r\star}}$ and the angle $\epsilon$ between \vec{V_{\mathrm{r\star}}} and the particle total velocity \vec{V}, so that

\begin{equation}
V_{\mathrm{r\oplus}} = V_{\mathrm{r\star}} \, \left( \cos{\theta} -\sin{\theta} \, \tan{\epsilon} \right).
\label{eq:vobs}
\end{equation}
\\
The layout of velocities and angles is displayed in Fig~\ref{schema}. The velocity observed from the Earth is more precisely $V_{\mathrm{r\oplus}} \, \sin{i}$, with \textit{i} the orbital inclination of the planet. Because of the high inclination for both planets (see Table~\ref{num_param}), Equation~\ref{eq:vobs} gives a good approximation of the observed velocity.

\subsubsection{Evolution with time}
\label{evol_time}

\begin{figure}[tbh]
\includegraphics[angle=-90,width=\columnwidth]{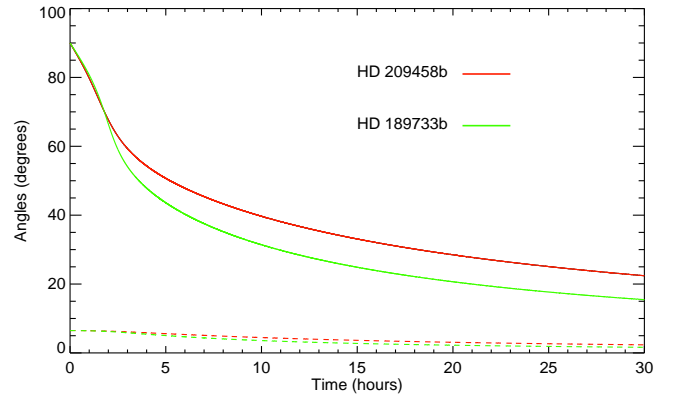}	
\caption[]{Angles $\epsilon$ (solid lines) between a particle radial velocity \vec{V_{\mathrm{r\star}}} and its total velocity \vec{V}, and $\theta_{\mathrm{lim}}$ (dashed lines), the maximum angular position of the particle at its ingress or egress. The angles are plotted as a function of time, counted from the escape of a particle from HD\,209458b (red line) or HD\,189733b (green line). The evolution of the angles is the same for both planets: they decrease with time, and $\epsilon$ is always greater than $\theta_{\mathrm{lim}}$. The slope is initially steeper for HD\,209458b because of its higher values for $\beta$ at low velocities, but the trend quickly reverses since $\beta$ decreases more rapidly with increasing velocity in the wings of the Lyman-$\alpha$ line of HD\,209458b (see Figs.~\ref{fit_HD189} and \ref{fit_HD209}).} 
\label{angles}
\end{figure}

\begin{figure}[tbh]
\includegraphics[angle=-90,width=\columnwidth]{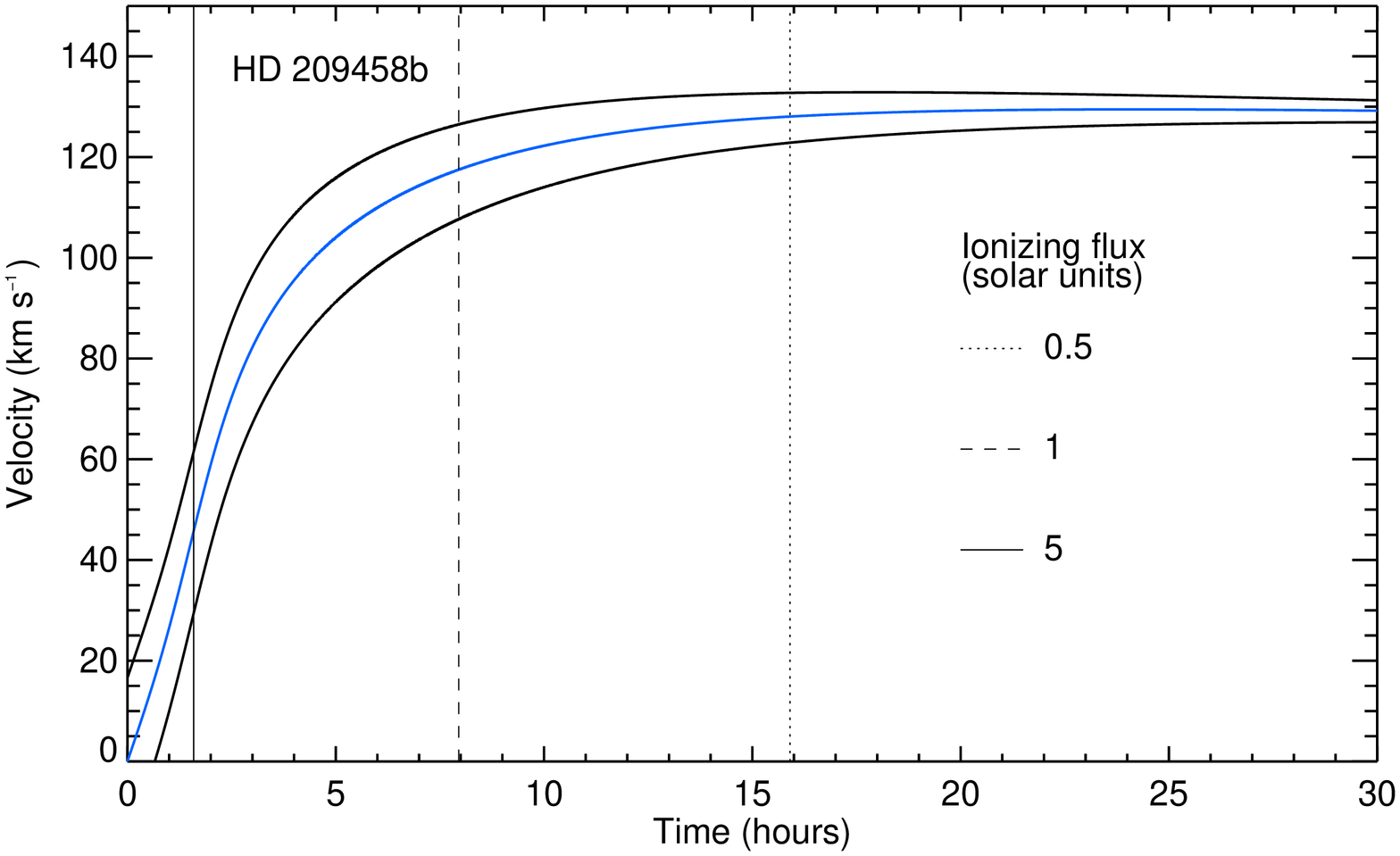}
\caption[]{Radial velocity (blue line), maximum observed velocity (black upper line) and minimum observed velocity (black lower line) as a function of time for a hydrogen atom escaped from HD\,209458b. The maximum velocity of $\sim$130\,km\,s$^{-1}$ is reached in about 10\,hours. The vertical black lines show the characteristic lifetimes of a hydrogen atom for stellar ionizing EUV fluxes of 0.5 (dotted line), 1 (dashed line) and 5 (solid line) times the solar value.} 
\label{rad_vel_HD209}
\end{figure}

\begin{figure}[tbh]
\includegraphics[angle=-90,width=\columnwidth]{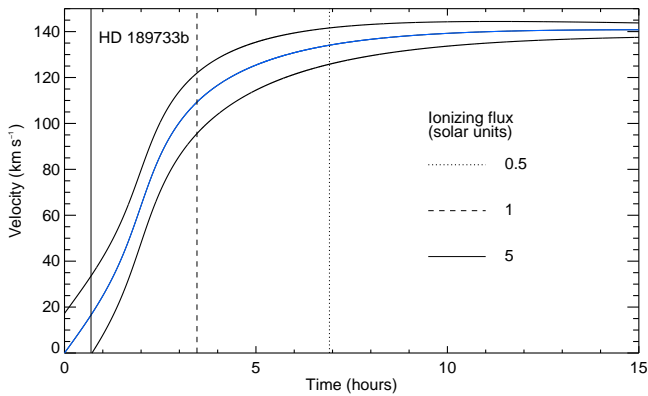}
\caption[]{Radial velocity (blue line), maximum observed velocity (black upper line) and minimum observed velocity (black lower line) as a function of time for a hydrogen atom launched from HD\,189733b. The maximum velocity of $\sim$140\,km\,s$^{-1}$ is reached in about 6\,hours. The vertical black lines show the characteristic lifetimes of a hydrogen atom for stellar ionizing EUV fluxes of 0.5 (dotted line), 1 (dashed line) and 5 (solid line) times the solar value. Because HD\,189733b is closer to its star than HD\,209458b, ionization lifetimes are shorter.} 
\label{rad_vel_HD189}
\end{figure}

We followed the trajectory and the velocity of the particle as a function of time since its escape from the planet's upper atmosphere (the particle 'proper time'). Since the particle initial velocity is that of the planet, its radial velocity is null and the radiation pressure coefficient is higher than unity (see Fig.~\ref{fit_HD189} and Fig.~\ref{fit_HD209}). The particle is then quickly accelerated to higher radial velocities, while the angle $\epsilon$ decreases (Fig.~\ref{angles}). Note that even if the hydrogen atoms are released with an additional thermal speed (as in the detailed 3D simulations), their initial velocity is still dominated by the planet orbital velocity. Using Equation~\ref{eq:vobs} we can estimate the velocity at which the particle would absorb the flux in the Lyman-$\alpha$ line observed from the Earth. However, we cannot know at what time during the transit observations we would observe the particle with this velocity, since it depends on when it escaped the planet. Instead, we searched for each proper time of the particle where it should be located in front of the stellar disk to be observed at its maximum or minimum velocity $V_{\mathrm{r\oplus}}$. The solution is quite straightforward, because the maximum velocity should be observed when $\theta=-\epsilon$, that is, when the total velocity of the particle is aligned with the line of sight toward the Earth. However, because a particle can only be observed from the Earth when it transits the stellar disk, its absolute angular position $\theta$ must remain lower than $\theta_{\mathrm{lim}}=\arctan{\frac{R_{*}}{r}}$ (with $R_{*}$ the star radius and $r$ the particle distance to the star, see Fig.~\ref{schema}). Besides, Fig.~\ref{angles} shows that $\epsilon$ is always higher than $\theta_{\mathrm{lim}}$. Since the observed velocity decreases when the particle angular position exceeds $-\epsilon$ and increases, its maximum value is $V_{\mathrm{r\oplus}}(\theta=-\theta_{\mathrm{lim}})$ at the particle ingress. Reciprocally, the minimum observed velocity is $V_{\mathrm{r\oplus}}(\theta=\theta_{\mathrm{lim}})$ at the particle egress. \\
The maximum and minimum observed velocities, as well as the radial velocity, are plotted as a function of time in Fig.~\ref{rad_vel_HD209} for HD\,209458b and Fig.~\ref{rad_vel_HD189} for HD\,189733b. First, we can see that a particle transiting the stellar disk can only absorb the Lyman-$\alpha$ line flux in a limited range dependent on its proper time. Then, all velocities increase with time until they reach their maximum values. This means that all the hydrogen atoms observed in front of the stellar disk will never absorb the flux in the Lyman-$\alpha$ line above a limit velocity if they are accelerated by radiation pressure only. For HD\,209458b, the theoretical maximum velocity of about 130\,km\,s$^{-1}$ is strikingly close to the upper limit of the absorption signature detected in 2001 (see Sect.~\ref{intro_HD209}). For HD\,189733b, the absorbing hydrogen atoms were observed above the radiation-pressure-limited velocity of about 140\,km\,s$^{-1}$ derived from Fig.~\ref{rad_vel_HD189}. An additional acceleration mechanism is needed to explain the absorption observed at such high velocities in the blue wing of the Lyman-$\alpha$ line.\\

\subsubsection{Ionization lifetime}
\label{ioniz}
The characteristic lifetime of a hydrogen atom subject to photoionization is inversely proportional to the stellar EUV ionizing flux (see Equation~\ref{eq:ion}). As a result, this flux has a strong influence on the depth and velocity structure of the absorption signatures from atoms accelerated by radiation pressure. Higher ionizing fluxes imply shorter ionization timescales and therefore fewer atoms at high velocity, since their lifetime is directly related to the velocity they can reach (Sect.~\ref{evol_time}). For example, an ionizing flux $F_{\mathrm{ion}}=1\,F_\odot$ associated to a characteristic lifetime of $\sim$8\,hours for HD\,209458b implies that about 15\% of the particles escaping the atmosphere will be accelerated to the maximum velocity of 130\,km\,s$^{-1}$ (Fig.~\ref{rad_vel_HD209}). When the ionizing flux increases, higher escape rates are needed to reach the same absorption depth at the same velocity.

%%%%%%%%%%%%%%%%%%%%%%%%%%%%%%%%%%%%%%%%%%%%%%%%%%%%%%%%%%%%%%%%%%%%%%%

\subsection{HD\,209458b: estimates of the hydrogen escape rate and the EUV ionizing flux}
\label{HD209}

\subsubsection{Best-fit parameters}

We used our 3D model to estimate the physical conditions in the exosphere of HD\,209458b needed to fit the absorption signature observed in the blue wing of the Lyman-$\alpha$ line in 2001. Note that with radiation pressure as the only acceleration mechanism, our model cannot reproduce the absorption signature that is potentially observed in the red wing of the line. However, the presence of this feature is far from being confirmed; it could be a false signal caused by statistical noise in contiguous pixels in the spectrum. The main free parameters of the model are the escape rate of neutral hydrogen from the planet's atmosphere and the EUV ionizing flux from the star. We calculated a grid of simulations as a function of these two parameters and compared the results with the observations using the $\chi^2$ defined by

\begin{equation}
\sum_{\mathrm{v}}{\left(\frac{A_{\mathrm{theo}}(v)-A_{\mathrm{obs}}(v)}{\sigma_{\mathrm{obs}}(v)}\right)^2},
\label{eq:ENA}
\end{equation}
where the fit is performed on the velocity interval described in Sect.~\ref{observations}. $A_{\mathrm{theo}}$ and $A_{\mathrm{obs}}$ are the theoretical and observed spectral absorption depths, with $\sigma_{\mathrm{obs}}$ the error on the observed absorption.\\

\begin{figure}[tbh]
\includegraphics[angle=-90,width=\columnwidth]{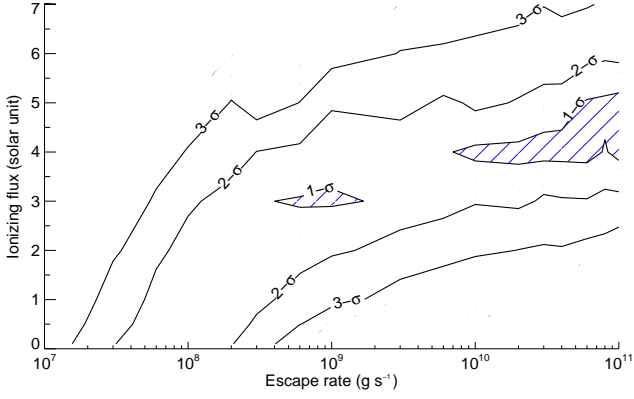}
\caption[]{Error bars for the estimated hydrogen escape rate and EUV ionizing flux for HD\,209458b. The best fits (at the 1$\sigma$ level, striped zones) are obtained in two regions: escape rate $\dot{M}\sim10^{9}$\,g\,s$^{-1}$ and ionizing flux $F_{\mathrm{ion}}\sim3\,F_\odot$, or escape rates above $6\times10^{9}$\,g\,s$^{-1}$ and ionizing fluxes $F_{\mathrm{ion}}\sim4\,F_\odot$. High escape rates are associated to high ionizing fluxes.} 
\label{chi2-HD209}
\end{figure}

\begin{figure}[tbh]
\includegraphics[angle=-90,width=\columnwidth]{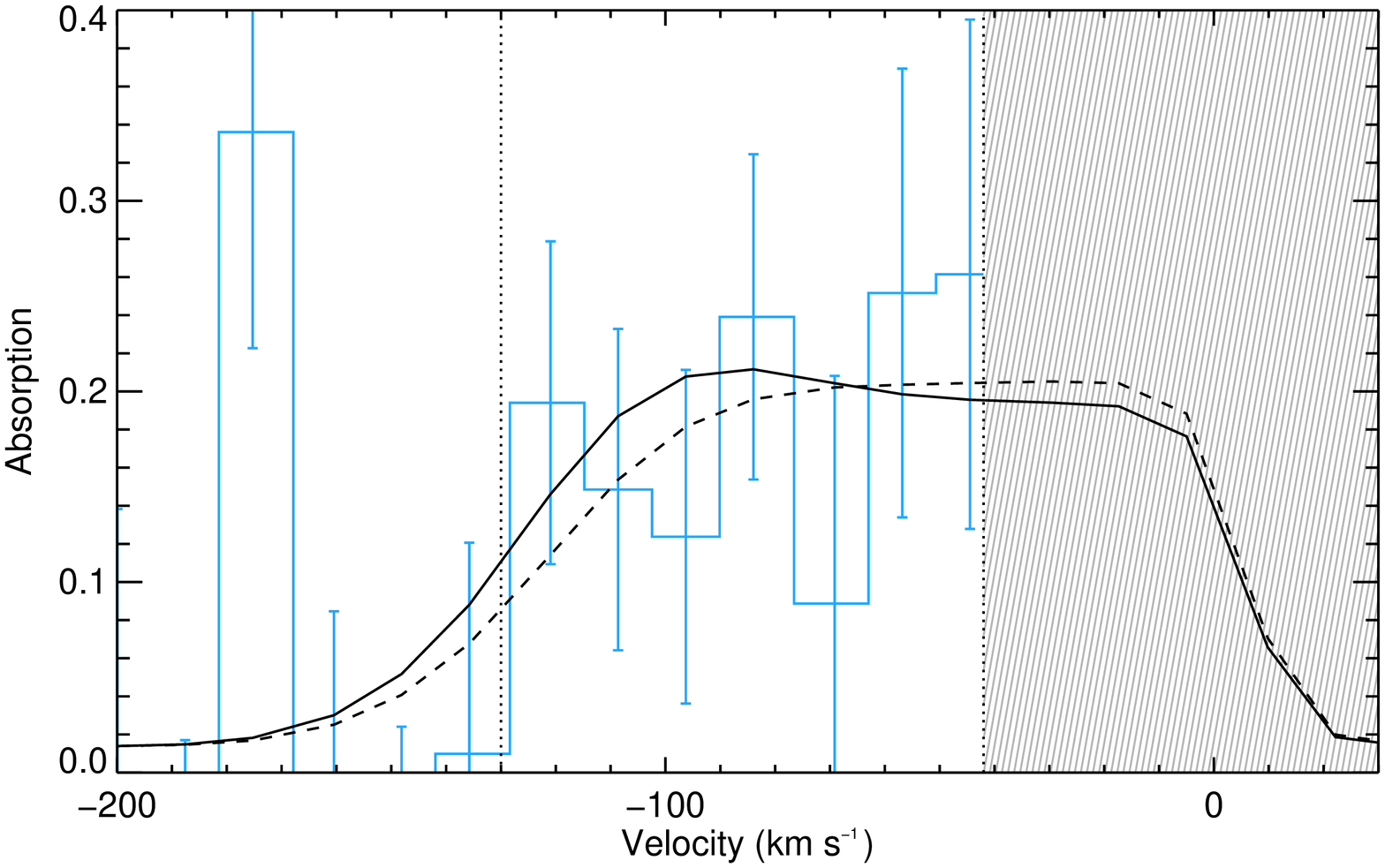}
\caption[]{Absorption profiles during the transit of HD\,209458b. The absorption signature observed in 2001 in the blue wing of the Lyman-$\alpha$ line is delimited by vertical black dotted lines (blue histogram; \citealt{VM2008}), and the striped gray zone corresponds to the region contaminated by the geocoronal emission. Two absorption profiles are calculated for best-fit parameters (solid line: escape rate \mbox{$\dot{M}=10^{9}$\,g\,s$^{-1}$} and ionizing flux $F_{\mathrm{ion}}=3\,F_\odot$; dashed line: escape rate \mbox{$\dot{M}=10^{10}$\,g\,s$^{-1}$} and ionizing flux $F_{\mathrm{ion}}=4\,F_\odot$). The simulated planetary disk is the source for an absorption depth of $\sim1.5\%$ at all wavelengths. A low escape rate results in a slightly decreased absorption depth at low velocity (solid line), while a high ionizing flux limits the number of atoms accelerated to high velocities (dashed line).} 
\label{best_fits_HD209}
\end{figure}

\begin{figure}[tbh]
\includegraphics[angle=-90,trim=0cm 0cm 8cm 0cm, clip=true,width=\columnwidth]{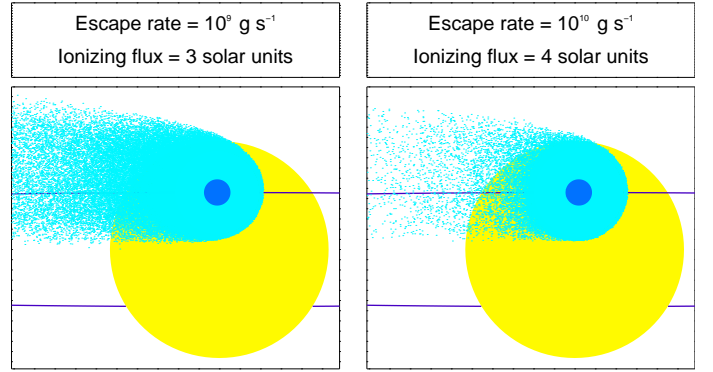}		
\caption[]{Best-fit simulations of the hot Jupiter HD\,209458b as seen along the star/Earth line of sight (parameters are given in the upper panels). Neutral hydrogen atoms (light blue dots) are escaping the planet (deep blue disk) at the center of the transit.} 
\label{best_fits_front_view}
\end{figure}

Two regions of the parameter space (Fig.~\ref{chi2-HD209}) were found to reproduce the observations whith a good $\chi^2$ of about 40 for 40 degrees of freedom. The first region corresponds to ionizing fluxes between 3.5 and 5\,$F_\odot$ and escape rates above $6\times10^{9}$\,g\,s$^{-1}$ (see also Figs.~\ref{best_fits_HD209} and \ref{best_fits_front_view}). This is consistent with mass loss of neutral hydrogen estimated to be in the range $10^{10}$ -- $10^{11}$\,g\,s$^{-1}$, either with a direct comparison with the observations ({\it e.g.}, \citealt{VM2003}; \citealt{Ehrenreich2008}) or with theoretical models ({\it e.g.}, \citealt{Lecav2004}; \citealt{Yelle2004}; \citealt{Tian2005}). Like \citet{Guo2013}, we found that the observations can also be explained by lower escape rates, in particular in a second region around $10^{9}$\,g\,s$^{-1}$ and about 3\,$F_\odot$. As explained in Sect.~\ref{ioniz}, with a higher ionizing flux, a higher escape rate is needed to reproduce the observations, up to the point that the lifetime of escaping atoms is so short that they cannot be accelerated to observable velocities by the radiation pressure (above 7\,$F_\odot$ at 3$\sigma$ from the best fit). On the other hand, for escape rates below $10^{8}$\,g\,s$^{-1}$ the hydrogen density is too low to significantly absorb the Lyman-$\alpha$ flux in the observed velocity range.\\

In contrast to the models of \citet{BJ_Hosseini2010} and \citet{Koskinen2010}, we cannot reproduce the velocity structure of the absorption profile by the sole natural and thermal broadening of the Lyman-$\alpha$ line (Sect.~\ref{abs_model}), and absorption arising mainly from hydrogen in the Roche lobe. Energetic neutral atoms have also been proposed as the source for the absorption signatures of HD\,209458b (\citealt{Holmstrom2008}; \citealt{Ekenback2010}; \citealt{Tremblin2012}). Although interactions with stellar wind protons cannot be excluded, our results show that radiation pressure alone, with its strength directly constrained by the observations, is enough to explain the data, without the need for additional acceleration mechanisms and additional degrees of freedom in the model. One can overlook radiation pressure by assuming a low, velocity-independent value for the coefficient $\beta$ ({\it e.g.}, \citealt{Ekenback2010} used a numerical input UV scattering rate of 0.35$\,s^{-1}$ that corresponds to a constant $\beta$=0.38) or considering that the atoms are ionized before they gain enough velocity ({\it e.g.}, \citealt{Koskinen2010}). We stress that radiation pressure is not a free parameter, but is derived from the Lyman-$\alpha$ line profile, which is in turn well reconstructed from real data (Sect.~\ref{rad_press}). In addition, even with short lifetimes of about 2\,hours (ionizing flux $F_{\mathrm{ion}}\sim5\,F_\odot$), enough atoms are accelerated to the observed velocities when the escape rate is high enough (see Fig.~\ref{chi2-HD209}).

\subsubsection{Cometary tail}
\label{com_tail}

\begin{figure}[tbh]		
\includegraphics[angle=-90,width=\columnwidth]{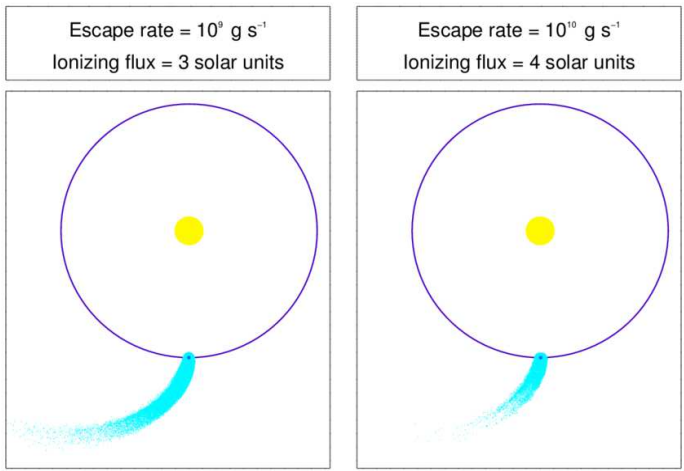}	
\caption[]{Views from the above of the planetary system of the hot Jupiter HD\,209458b, for best-fit simulations (parameters are given in the upper panels). Neutral hydrogen atoms (light blue dots) escaping the planet (deep blue disk) organize into a cometary tail. A lower ionizing flux results in a longer and denser tail, composed of atoms accelerated to high velocities. Both simulations fit well the observations because of the loose constraints on the post-transit.} 
\label{sim_views}
\end{figure}

As was expected from the physics of our model and previous analyses ({\it e.g.}, \citealt{VM_lecav2004}; \citealt{Schneiter2007}; \citealt{Ehrenreich2008}), the simulations show the formation of a hydrogen cometary tail trailing behind the planet (Fig.~\ref{sim_views}). As a result, the transit of escaping hydrogen may last longer than the optical occultation by the planetary disk alone. Since the ionizing flux limits the number of atoms that reach a given velocity, it also constrains the spatial extension of the cloud as well as the depth and duration of the transit and post-transit occultations (Sect.~\ref{spec_temp}). For HD\,209458b, the lack of measurements after the transit in 2001 allows the observations to be reproduced with a shortened cometary tail (Figs.~\ref{best_fits_front_view} and \ref{sim_views}), nearly completely ionized for an ionizing flux $F_{\mathrm{ion}}\sim5\,F_\odot$. Note, however, that our simulations do not support the results that the observed absorption profile may have a symmetric shape (see for example \citealt{BJ_Hosseini2010}), as can be seen on Fig.~\ref{best_fits_HD209}. The ACS post-transit observation from \citet{Ehrenreich2008} constrains our results loosely because of the large error bar on the data. Nonetheless, the non-detection of absorption after the transit requires a shorter cometary tail and shifts the inferior levels at 2 and 3$\sigma$ toward higher ionizing fluxes in Fig.~\ref{chi2-HD209}.

\subsubsection{Influence of the self-shielding}
\label{self_views}
\begin{figure*}
\centering
\begin{minipage}[b]{0.9\textwidth}	
\includegraphics[angle=-90,width=\columnwidth]{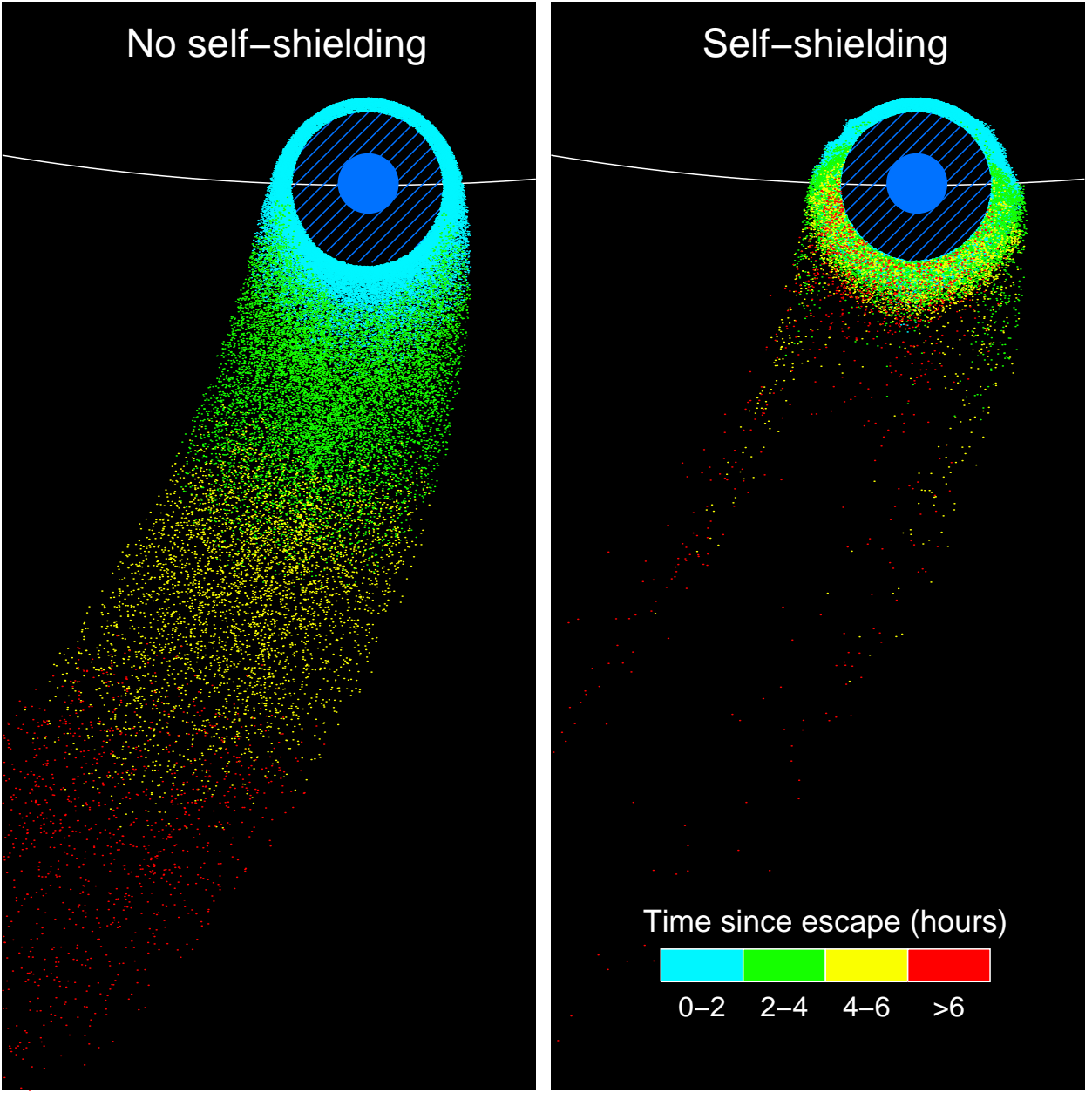}
\end{minipage}
\caption[]{Distribution of hydrogen atoms in the orbital plane of HD\,209458b using best-fit parameters (escape rate \mbox{$\dot{M}=10^{10}$\,g\,s$^{-1}$}; ionizing flux $F_{\mathrm{ion}}=4\,F_\odot$). The color scale indicates the time since the atoms escaped the upper atmosphere. Without self-shielding (left panel), the atoms are homogeneously accelerated away from the star by radiation pressure. With self-shielding (right panel), the internal regions of the cloud are submitted to a lower radiation pressure and are accelerated on longer timescales. The atmosphere below the exobase is uniformely filled with high-density neutral hydrogen gas (striped disk centered on the planet in deep blue).}
\label{orbital_plane}
\end{figure*}

\begin{figure}[tbh]
\includegraphics[angle=-90,width=\columnwidth]{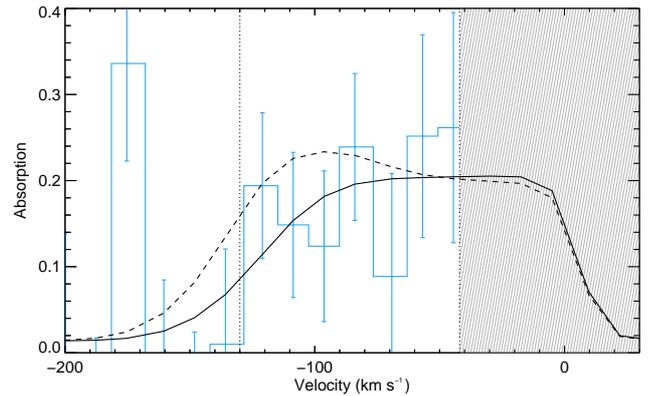}
\caption[]{Absorption profiles during the transit of HD\,209458b. The absorption signature observed in 2001 in the blue wing of the Lyman-$\alpha$ line is delimited by vertical black dotted lines (blue histogram; \citealt{VM2008}), and the striped gray zone corresponds to the region contaminated by the geocoronal emission. Two absorption profiles are calculated for the same best-fit parameters (escape rate $\dot{M}=10^{10}$\,g\,s$^{-1}$; ionizing flux $F_{\mathrm{ion}}=4\,F_\odot$), one with self-shielding (black solid line) and the other without (black dashed line). Taking into account self-shielding decreases the absorption depth at high velocities.} 
\label{abs_self_noself}
\end{figure}

The high-density regions of the cloud near the planet are strongly shielded from stellar photons. As a result, a great number of atoms is subject to lower radiation pressure and remain at low radial velocities much longer than was expected from the simplified 2D simulations detailed in Sect.~\ref{hyd_dyn}. This effect can be observed more clearly in the planet orbital plane (Fig.~\ref{orbital_plane}). Even if the bulk of the planetary wind is shielded from radiative blow-out, as estimated by \citet{Tremblin2012}, our results show that the escape of hydrogen at the limbs of the atmosphere is enough to reproduce the observations, provided the atoms are moving at high velocities in the blue wing (all simulated absorption profiles at less than $1\sigma$ from the best fit have a velocity structure similar to that of Fig.~\ref{best_fits_HD209}), which is easily explained by radiation pressure acceleration.\\
Self-shielding from photoionization is negliglible (Sect.~\ref{self-shielding}). Thus, even if hydrogen atoms shielded from radiation pressure are accelerated on longer timescales, they are still ionized with the same characteristic lifetimes. As a consequence, many of these atoms are ionized before they are accelerated to the velocities they would have reached without self-shielding. The absorption profile is consequently shallower at high velocities with self-shielding, but is not as affected at low velocities where the absorption appears to be saturated because of the optically thick regions of the cloud close to the planet (Fig.~\ref{abs_self_noself}).

%%%%%%%%%%%%%%%%%%%%%%%%%%%%%%%%%%%%%%%%%%%%%%%%%%%%%%%%%%%%%%%%%%%%%%%

\subsection{HD\,189733b: evaporating state in September 2011}
\label{HD189}

For HD\,189733b, the high velocities of the absorption signature detected in the blue wing of the line in 2011 can be explained if the atoms are accelerated, in addition to the radiation pressure, by interactions with stellar wind protons (Sect.~\ref{ENA}). We fitted the observations using a model with five free parameters: the neutral hydrogen escape rate, the stellar ionizing flux, and the stellar wind properties (density at the orbit of the planet, bulk velocity, and temperature). We found that the bulk velocity is tightly constrained to $V_{\mathrm{wind}}=200\pm20$\,km\,s$^{-1}$, close to the velocity of $190$\,km\,s$^{-1}$ estimated by \citet{Lecav2012}. This is in line with a slow stellar wind, as emitted by the Sun in its equatorial regions ({\it e.g.}, \citealt{McComas2003}; \citealt{Quemerais2007}), and consistent with the small spin-orbit misalignment of HD\,189733b (\mbox{$\beta=0.85^\circ$$\stackrel{0.32}{_{-0.28}}$}, see \citealt{Triaud2009}). \citet{Ekenback2010} and \citet{Tremblin2012} used $V_{\mathrm{wind}}=450$\,km\,s$^{-1}$ and $V_{\mathrm{wind}}=130$\,km\,s$^{-1}$ in their respective modeling of the stellar wind for HD\,209458. In the following discussion, the bulk velocity was set to $V_{\mathrm{wind}}=200$\,km\,s$^{-1}$.
The temperature of the proton distribution controls the spread of the absorption profile by high-velocity neutral atoms (\citealt{Holmstrom2008}). Higher temperatures increase the width of the velocity range of the absorption profile, and decrease its depth (conversely for lower temperatures). We found that a temperature $T_{\mathrm{wind}}=3\times10^{4}\,K$ fits the observations well and used it for all simulations hereafter. This temperature is about ten times lower than solar-scaled values determined by, for example, \citet{Lemaire2011}. 
Assuming their tentative detection of an early-ingress absorption in the C\,{\sc ii} line of HD\,189733b is caused by the planet's magnetosphere, constrained by the stellar wind, \citet{BJ_Ballester2013} derived rough limitations on the wind velocity ($V_{\mathrm{wind}}=200$ to 900\,km\,s$^{-1}$) and temperature ($T_{\mathrm{wind}}<4\times10^{6}\,K$).\\
We note that in our model, no absorption is simulated in the red wing of the Lyman-$\alpha$ line, whereas a hint of absorption was observed in this part of the spectrum in 2011 (\citealt{Lecav2012}). However, as for HD\,209458b, this feature is not confirmed, and the probability that it is a false signal caused by statistical noise in the spectrum has been estimated to be 25\% (\citealt{Bourrier2013}).

\subsubsection{Escape-limited saturation regime}
\label{saturation}

\begin{figure}[tbh]
\includegraphics[angle=-90,width=\columnwidth]{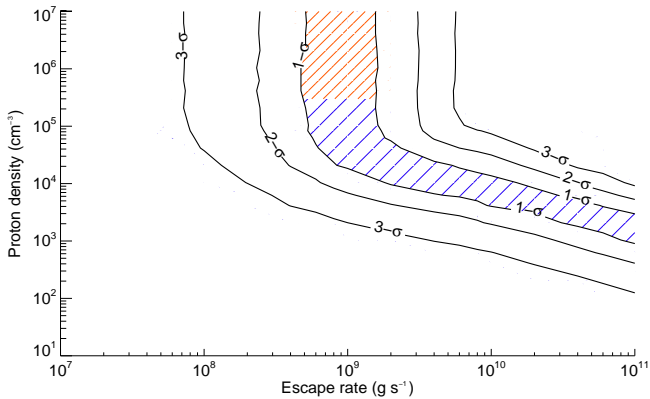}
\caption[]{Error bars on the estimated atmospheric hydrogen escape rate and stellar proton density for HD\,189733b. The ionizing flux is fixed to $10\,F_\odot$. Observations are best fitted inside the striped zones, with a $\chi^2$ of $62.8\pm1$ for 87 degrees of freedom. Proton densities above $\sim3\times10^{5}$\,cm$^{-3}$ (orange striped zone) correspond to an 'escape-limited' saturation regime in which all escaping hydrogen atoms interact with the stellar wind. Outside of this regime, observations are well reproduced for decreasing proton densities with higher escape rates.} 
\label{chi2_HD189_flux10}
\end{figure}

After constraining the temperature and velocity of the proton stellar wind, we found that the model best fits the observations with an ionizing flux $F_{\mathrm{ion}}=10\,F_\odot$. The $\chi^2$ is calculated as in Sect.~\ref{HD209}. The same good value of $\sim63$ for 87 degrees of freedom is found for a wide range of hydrogen escape rates and proton density. Fig.~\ref{chi2_HD189_flux10} shows the goodness of fit as a function of the escape rate and proton density for an ionizing flux $F_{\mathrm{ion}}=10\,F_\odot$. A remarkable feature of this diagram is that for any escape rate in the range $5\times10^{8}$ to $1.5\times10^{9}$\,g\,s$^{-1}$, an increase in the proton density above $\sim3\times10^{5}$\,cm$^{-3}$ has no influence on the quality of the fit, as shown by the constant $\chi^2$ of about 63 at the 1$\sigma$ level. It corresponds to a \textit{saturation regime}, 'escape-limited', in which most neutral hydrogen atoms escaping the atmosphere are accelerated by the stellar wind to high velocities. Accordingly, there is no atmospheric absorption at low velocities in this regime. \\

With lower proton densities, outside of the saturation regime, there are escaping atoms that are accelerated by radiation pressure without being subject to charge-exchange before they are ionized by EUV photons. As a result, fewer atoms are accelerated to the high velocities of the absorption signature and higher escape rates are needed to fit the observations (Fig.~\ref{chi2_HD189_flux10}). The observations can be well reproduced for proton densities between $10^{3}$ and $3\times10^{5}$\,cm$^{-3}$ and escape rates above $5\times10^{8}$\,g\,s$^{-1}$. This is consistent with the upper limit on the proton density estimated by \citet{BJ_Ballester2013} to be $5\times10^{7}$\,cm$^{-3}$. A solar-like wind would correspond to a proton density of about $7\times10^{3}$\,cm$^{-3}$ at the orbit of HD\,189733b (\citealt{Quemerais2007}). Note that for escape rates higher than $10^{11}$\,g\,s$^{-1}$, there can be another saturation regime, 'wind-limited', in which all stellar protons interact with the neutral hydrogen gas. In this regime we would expect to detect a large fraction of hydrogen atoms accelerated only by radiation pressure in the velocity range -140 to 0\,km\,s$^{-1}$. However, we see no such population of atoms in the observations of HD\,189733b, and the escaping atmosphere is not likely in this regime.

\begin{figure}[tbh]
\includegraphics[angle=-90,width=\columnwidth]{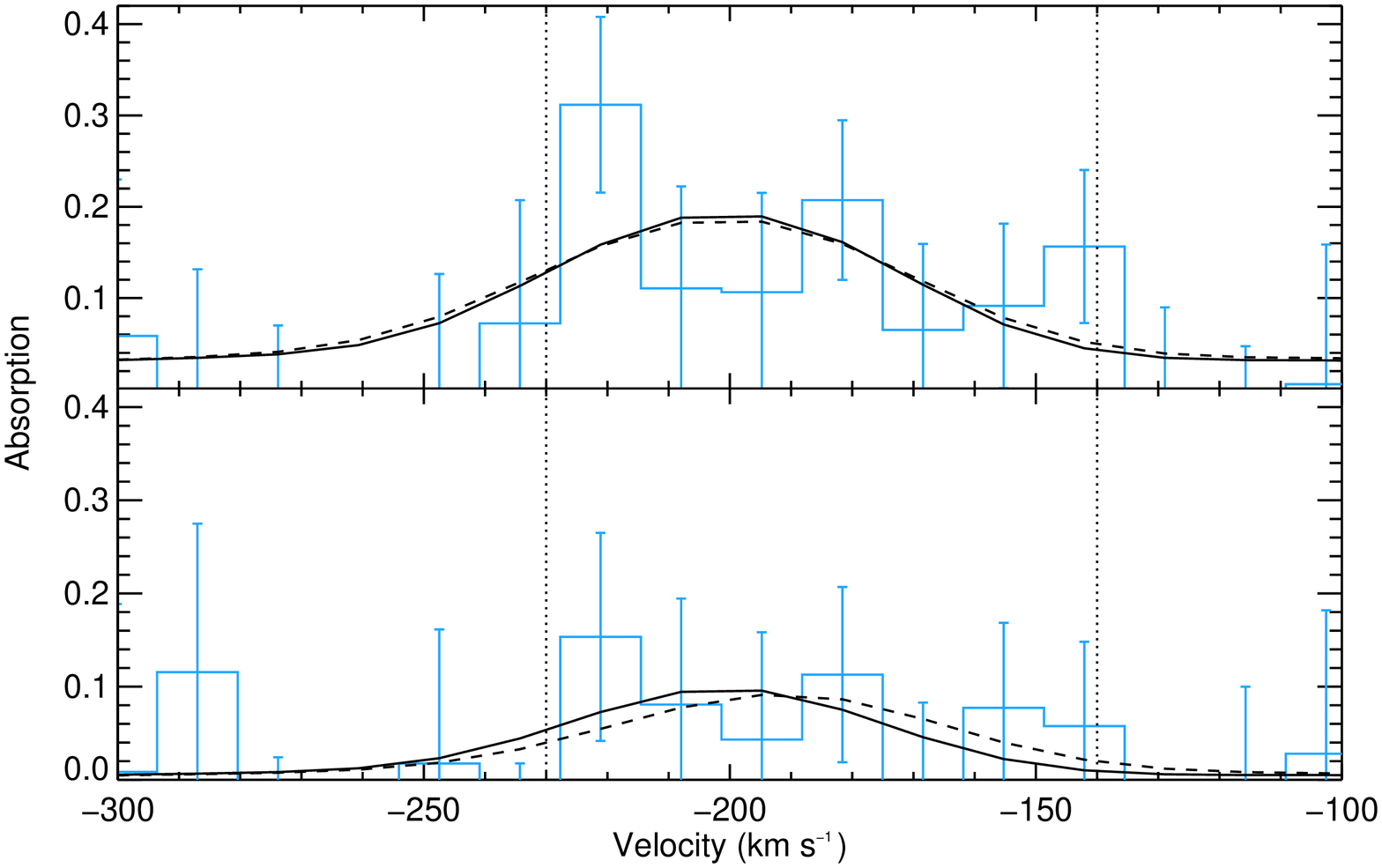}
\caption[]{Absorption profiles during the transit (top panel) and the post-transit (bottom panel) of HD\,189733b. The absorption signature observed in 2011 at high velocities in the blue wing of the Lyman-$\alpha$ line is delimited by vertical black dotted lines (blue histogram; \citealt{Lecav2012} and \citealt{Bourrier2013}). Two absorption profiles are calculated for best-fit parameters (ionizing flux $F_{\mathrm{ion}}=10\,F_\odot$) in the saturation regime (solid line: escape rate $\dot{M}=10^{9}$\,g\,s$^{-1}$ and proton density $n_{\mathrm{wind}}=10^{6}$\,cm$^{-3}$) and out of this regime (dashed line: escape rate $\dot{M}=8\times10^{10}$\,g\,s$^{-1}$ and proton density $n_{\mathrm{wind}}=2\times10^{3}$\,cm$^{-3}$). The main difference between the two profiles is the absorption at the center of the line, which is significantly reduced in the saturation regime, although for HD\,189733b the Lyman-$\alpha$ flux at these wavelengths is in any case completely absorbed by the blueshifted ISM (see Fig.~\ref{fit_HD189}). The simulated planetary disk causes an absorption depth of $\sim2.3\%$ at all wavelengths during the transit.} 
\label{best_fits_HD189}
\end{figure}

\subsubsection{Impact of the ionizing flux}
\label{ion_2011}

\begin{figure}[tbh]
\includegraphics[angle=-90,width=\columnwidth]{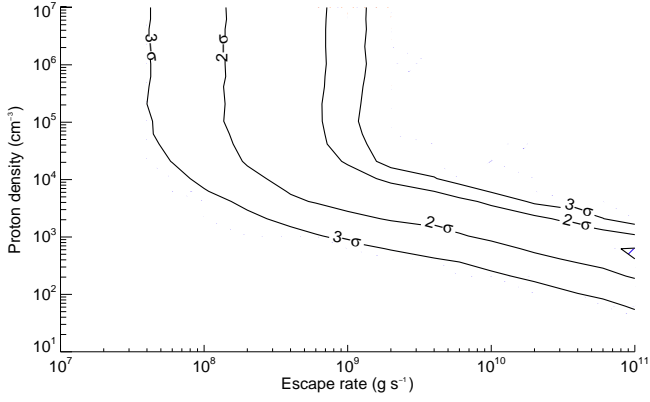}
\caption[]{Same plot as in Fig.~\ref{chi2_HD189_flux10} for an ionizing flux \mbox{$F_{\mathrm{ion}}=5\,F_\odot$}. Values fitting the observations are shifted toward lower escape rates and proton densities. There are nearly no values left that fit the observations at less than $1\sigma$ from the best fit obtained for $10\,F_\odot$.} 
\label{chi2_HD189_flux5}
\end{figure}

\begin{figure}[tbh]
\includegraphics[angle=-90,width=\columnwidth]{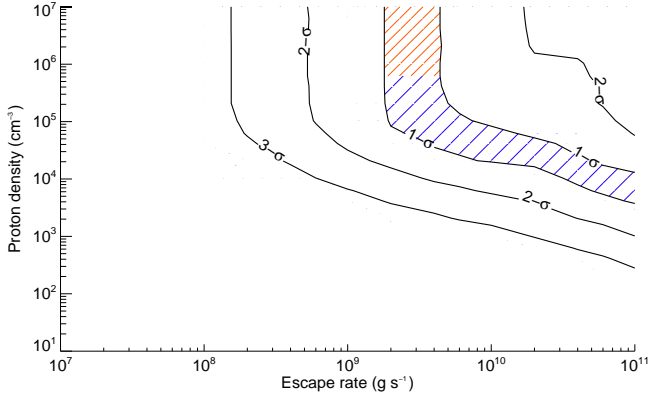}
\caption[]{Same plot as in Fig.~\ref{chi2_HD189_flux10} for an ionizing flux \mbox{$F_{\mathrm{ion}}=20\,F_\odot$}. Values fitting the observations are shifted toward higher escape rates and proton densities.} 
\label{chi2_HD189_flux20}
\end{figure}

We studied the influence of the ionizing flux by making the same analysis as in Sect.~\ref{saturation} for $F_{\mathrm{ion}}=5\,F_\odot$ (Fig.~\ref{chi2_HD189_flux5}) and $F_{\mathrm{ion}}=20\,F_\odot$ (Fig.~\ref{chi2_HD189_flux20}). Our results are consistent with the previous estimation made by \citet{Lecav2012} for the 2011 observations, with an escape rate $\dot{M}=10^{9}$\,g\,s$^{-1}$ and a proton density $n_{\mathrm{wind}}=3\times10^{3}$\,cm$^{-3}$ for an ionizing flux $F_{\mathrm{ion}}=5\,F_\odot$. With higher ionizing fluxes, higher escape rates and proton densities are needed to fit the observations. Indeed, the interactions of the stellar wind with the escaping hydrogen gas create neutral atoms moving instantaneously at the Maxwellian velocities of the proton distribution, without the need for progressive acceleration by radiation pressure (Sect.~\ref{ENA}). These atoms responsible for the absorption at high velocities are thus produced anew at every moment, and the ionization has less effect on their distribution than it has on the atoms escaping the atmosphere. However, higher ionizing fluxes result in fewer escaping atoms able to interact with the stellar wind before they are ionized. The ensuing depletion of high-velocity atoms created by these interactions can be compensated for, up to a certain extent, by increasing the density of either the stellar or the planetary wind (i.e., the escape rate). Note that the ionizing flux has to be high enough for atoms escaping the atmosphere without interacting with the stellar wind to be ionized before they contribute to the absorption over the interval -140 to 0\,km\,s$^{-1}$, where no signature was detected in 2011. The limit on the proton density to enter the escape-limited saturation regime increases slightly with the ionizing flux (from $n_{\mathrm{wind}}=3\times10^{5}$\,cm$^{-3}$ for $F_{\mathrm{ion}}=10\,F_\odot$, to $n_{\mathrm{wind}}=6\times10^{5}$\,cm$^{-3}$ for $F_{\mathrm{ion}}=20\,F_\odot$).\\

\begin{figure}[tbh]
\includegraphics[angle=-90,width=\columnwidth]{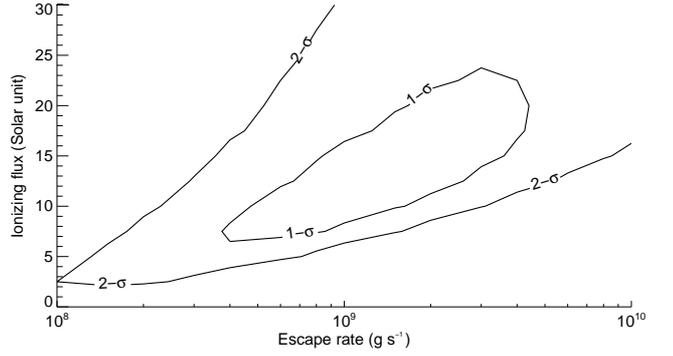}
\caption[]{Error bars at the 1 and $2\sigma$ levels for the hydrogen escape rate and ionizing flux in the saturation regime of HD\,189733b. This plot is independent of the proton density for values above $\sim6\times10^{5}$\,cm$^{-3}$.} 
\label{chi2_HD189_flux}
\end{figure}

We found that the span of the 'valley' fitting the observations at less than $1\sigma$ from the best fit (striped zone in Fig.~\ref{chi2_HD189_flux10}, \ref{chi2_HD189_flux5} and \ref{chi2_HD189_flux20}) depends strongly on the ionizing flux. The broadest ranges of escape rates and proton densities that reproduce the observed absorption profile are obtained for an ionizing flux $F_{\mathrm{ion}}\sim10\,F_\odot$. Decreasing \textit{or} increasing the ionization shrinks the valley. The saturation regime zone has more or less the broadest span of the valley, and so we used it as a way to estimate error bars on the ionizing flux. Since this regime does not depend on the proton density, we studied the relation between escape rate and ionizing flux for a fixed proton density $n_{\mathrm{wind}}=10^{6}$\,cm$^{-3}$ (Fig.~\ref{chi2_HD189_flux}). The saturation regime, characterized by a lower limit on the proton density in the range $3\times10^{5}$ -- $6\times10^{5}$\,cm$^{-3}$, can be furthermore defined at the $1\sigma$ level by escape rates in the range \mbox{$4\times10^{8}$ -- $4\times10^{9}$\,g\,s$^{-1}$} and ionizing fluxes in the range \mbox{6 -- 23$\,F_\odot$}. Note that an unresolved observation of Lyman-$\alpha$ absorption made in 2007-2008 with ACS was fitted at the $1\sigma$ level by \citet{Lecav2010}. Taking into account radiation pressure only, they found escape rates in the range $3\times10^{9}$ -- $10^{11}$\,g\,s$^{-1}$ and ionizing fluxes in the range 15 -- 40\,$F_\odot$.

\subsection{HD\,189733b: non-detection in April 2010}
\label{noevap}

While in September 2011 neutral hydrogen was detected in the upper atmosphere of HD\,189733b through absorption in the Lyman-$\alpha$ line, no absorption was observed in April 2010 (\citealt{Lecav2012}). After constraining the physical conditions consistent with the 2011 observations (Sect.~\ref{HD189}), we now determine the variations of the escape rate, the ionizing flux, or the stellar wind density needed to explain the changes observed between 2010 and 2011. Using a configuration without any atmospheric escape as reference (this provides the best-fit to the 2010 data), we estimate limits on these parameters beyond which the absorption by the escaping neutral hydrogen would have been detected in the 2010 observations.

\subsubsection{Variations of the escape rate and ionizing flux}

\begin{figure}[tbh]
\includegraphics[angle=-90,width=\columnwidth]{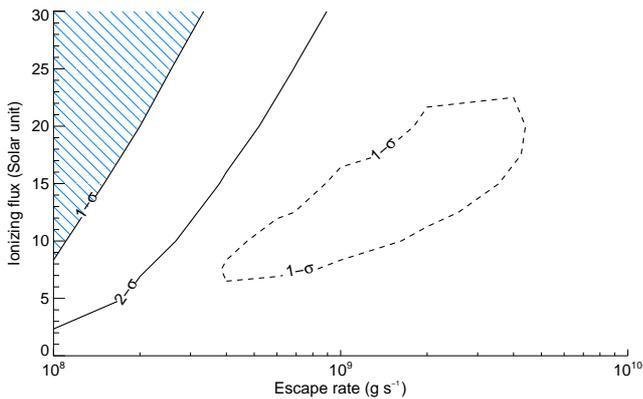}
\caption[]{Detection levels of Lyman-$\alpha$ absorption in the 2010 data of HD\,189733b as a function of the hydrogen escape rate and the ionizing flux (thick black lines), assuming that the stellar wind is in the 'escape-limited' saturation regime.
Beyond the striped blue zone, absorption would have been detected at more than $1\sigma$. For comparison, the dashed black line shows the error bars on the best fit to the absorption signature observed in 2011 (see Fig.~\ref{chi2_HD189_flux}).}
\label{chi2_HD189_flux_2010_wind}
\end{figure}

First, we considered no change in the stellar wind, with the same wind properties that provided a best-fit to the 2011 observations ($V_{\mathrm{wind}}=200$\,km\,s$^{-1}$, $T_{\mathrm{wind}}=3\times10^{4}\,K$ and $n_{\mathrm{wind}}=10^{6}$\,cm$^{-3}$). For the same ionizing fluxes as in 2011 (6 -- 23$\,F_\odot$), the non-detection of absorption in Lyman-$\alpha$ in 2010 can be explained at the 1-$\sigma$ level by escape rates that are about 5 to 20 times lower (Fig.~\ref{chi2_HD189_flux_2010_wind}). Indeed, with fewer neutral hydrogen atoms escaping the planet atmosphere, the population of high-velocity atoms is so reduced that their absorption signature in the Lyman-$\alpha$ line is below the detection level of the 2010 data. In contrast, no realistic increase of the stellar ionizing flux can explain the temporal variations observed in Lyman-$\alpha$. Indeed, if the escape rate were the same as in 2011 ($4\times10^{8}$ -- $4\times10^{9}$\,g\,s$^{-1}$), the non-detection in 2010 would require an ionizing flux higher than at least 30 times the solar value (Fig.~\ref{chi2_HD189_flux_2010_wind}).

Note that the stellar wind considered above corresponds to the 'escape-limited' saturation regime, in which all neutral hydrogen atoms that escape the atmosphere interact with the stellar wind (see Sect.~\ref{saturation}). Nonetheless, we found that for any stellar wind properties and ionizing fluxes consistent with the 2011 detection, the 2010 observation can always be explained by a decrease of the escape rate by about an order of magnitude.

\subsubsection{Variations in the stellar wind}

\begin{figure}[tbh]
\includegraphics[angle=-90,width=\columnwidth]{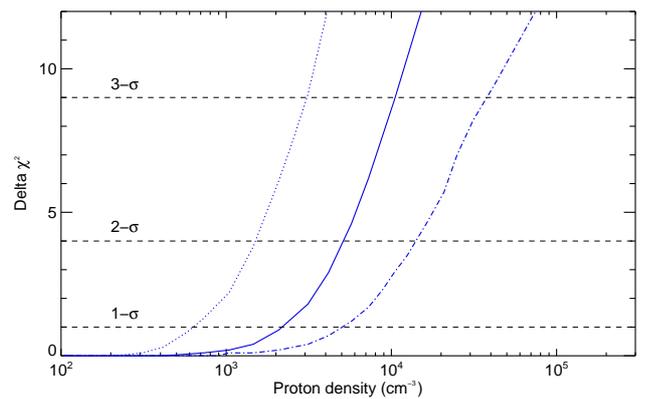}
\caption[]{$\chi^2$ difference between simulations of the 2010 data and the best fit without atmospheric escape as a function of the stellar wind density, assuming a constant escape rate and ionizing flux ($\dot{M}=2\times10^{9}$\,g\,s$^{-1}$ and $F_{\mathrm{ion}}=10\,F_\odot$, thick blue line). The intersection with the black dotted lines gives the proton density above which an absorption signature would have been detected at more than 1, 2 or 3 $\sigma$ in the 2010 observations. The upper limit on the proton density decreases with higher escape rates ($\dot{M}=2\times10^{10}$\,g\,s$^{-1}$ and $F_{\mathrm{ion}}=10\,F_\odot$, blue dotted line) and increases with higher ionizing fluxes ($\dot{M}=2\times10^{9}$\,g\,s$^{-1}$ and $F_{\mathrm{ion}}=20\,F_\odot$, blue dash-dotted line).} 
\label{chi2_HD189_flux_2010_nowind}
\end{figure}

Here we considered no change in the escape rate and the ionizing flux, with the same values that provided one of the best fits to the 2011 observations ($\dot{M}=2\times10^{9}$\,g\,s$^{-1}$ and $F_{\mathrm{ion}}=10\,F_\odot$). Thus we investigate whether variations in the stellar wind can explain the non-detection in 2010. We found that with a proton density below 2$\times$10$^{3}$\,cm$^{-3}$ (ten times lower than the best fit to the 2011 data with $n_{\mathrm{wind}}=2\times10^{4}$\,cm$^{-3}$) the resulting spectrum can be consistent within 1-$\sigma$ with the 2010 observations (Fig.~\ref{chi2_HD189_flux_2010_nowind}). With higher ionizing fluxes (but still consistent with the 2011 data) this upper limit on the stellar wind proton density is slightly higher (5$\times$10$^{3}$\,cm$^{-3}$ for $F_{\mathrm{ion}}=20\,F_\odot$), while higher escape rates provide lower upper limits on the wind density (6$\times$10$^{2}$\,cm$^{-3}$ for $\dot{M}=2\times10^{10}$\,g\,s$^{-1}$). This shows that the observed variations in the Lyman-$\alpha$ transit of HD\,189733b can be explained by variations in the wind properties only.

Note that with a low wind density, many neutral hydrogen atoms that escape the atmosphere do not interact with the stellar wind. Because they are still accelerated by radiation pressure, they produce absorption that is limited to the velocity range -140 to 0\,km\,s$^{-1}$. However, such absorption signatures cannot be detected in the 2010 data because of the interstellar absorption in this part of the spectrum and the wide airglow emission (up to -60\,km\,s$^{-1}$ in the blue wing of the Lyman-$\alpha$ line in April 2010). With the same escape rates as in 2011, the atoms accelerated by radiation-pressure could only have been detected if the ionization timescale had been longer, i.e., the ionizing flux had been at a very low value, which is incompatible with the 2011 observations (see Sect.~\ref{ion_2011}).\\

In conclusion, the changes observed in the upper atmosphere of HD\,189733b between April 2010 and September 2011 can be explained by two different scenarios. The observations can be reproduced with an increase of the escape rate by about an order of magnitude between 2010 and 2011. Alternatively, a reduced stellar wind can explain the non-detection of Lyman-$\alpha$ absorption in 2010.

%%%%%%%%%%%%%%%%%%%%%%%%%%%%%%%%%%%%%%%%%%%%%%%%%%%%%%%%%%%%%%%%%%%%%%%

\section{Short timescale spectro-temporal variability of the Lyman-$\alpha$ absorption spectra}
\label{spec_temp}

\subsection{Time variations}
\label{time_var}

\begin{figure*}
\centering
\begin{minipage}[b]{0.9\textwidth}	
\includegraphics[angle=-90,width=\columnwidth]{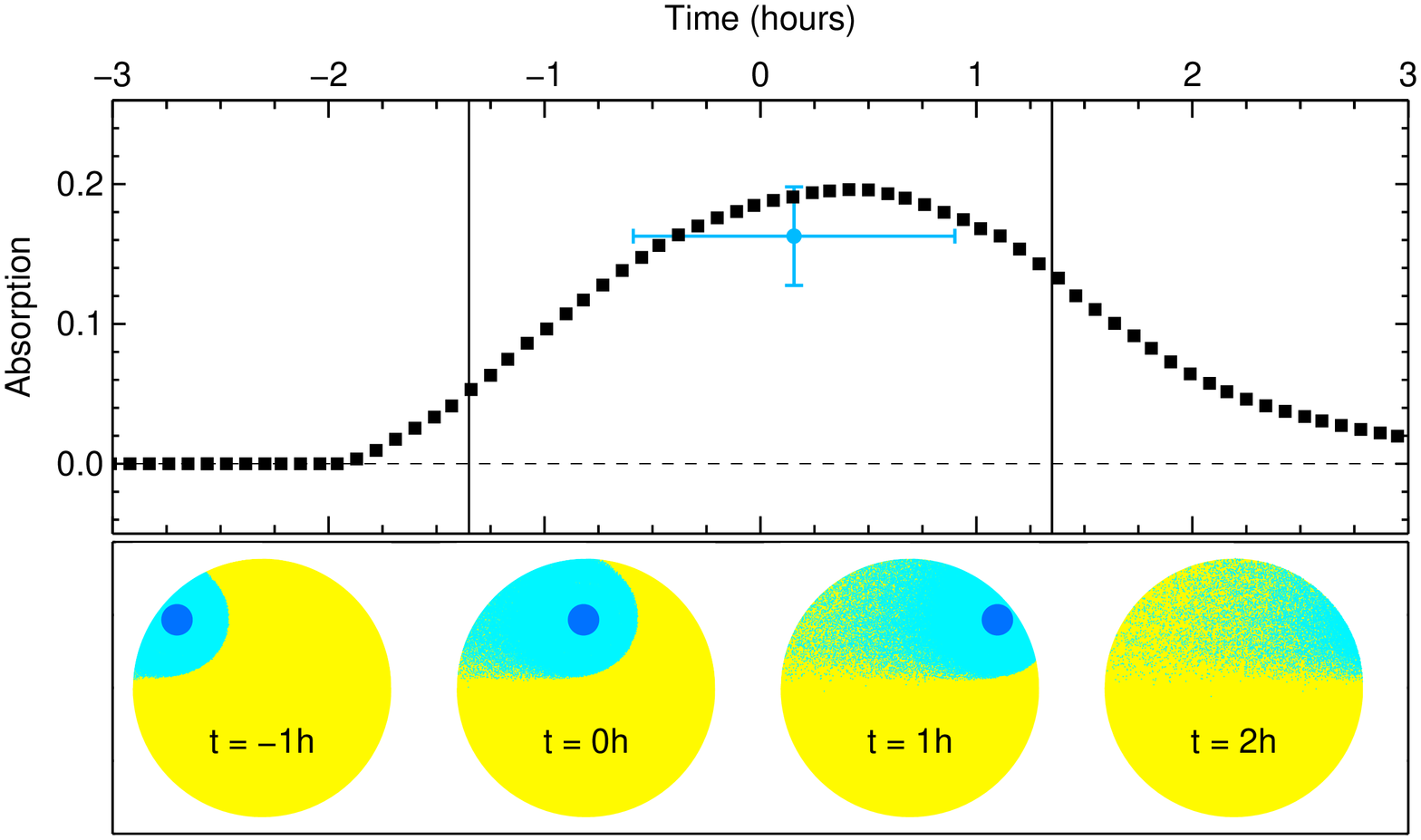}	
\caption[]{\textit{Top panel:} Total absorption over the range -130 to -40\,km\,s$^{-1}$, for a simulation of atmospheric escape from HD\,209458b (escape rate $\dot{M}=10^{9}$\,g\,s$^{-1}$; ionizing flux $F_{\mathrm{ion}}=3\,F_\odot$). Vertical solid lines show the beginning and end of ingress and egress of the transit. The blue point shows the observed absorption, with the horizontal error bar displaying the duration of the observation. Each square corresponds to a time step in the simulation. The variations of the absorption depth are correlated to the optical depth of the hydrogen cloud and the stellar surface it occults. \textit{Bottom panel:} Views of the gas (light blue) and planet (deep blue) transit along the star/Earth line of sight. Only particles with observed velocities in the range -130 to -40\,km\,s$^{-1}$ are represented.} 
\label{sim_tot_abs_HD209}
\end{minipage}
\begin{minipage}[b]{0.9\textwidth}	
\includegraphics[angle=-90,width=\columnwidth]{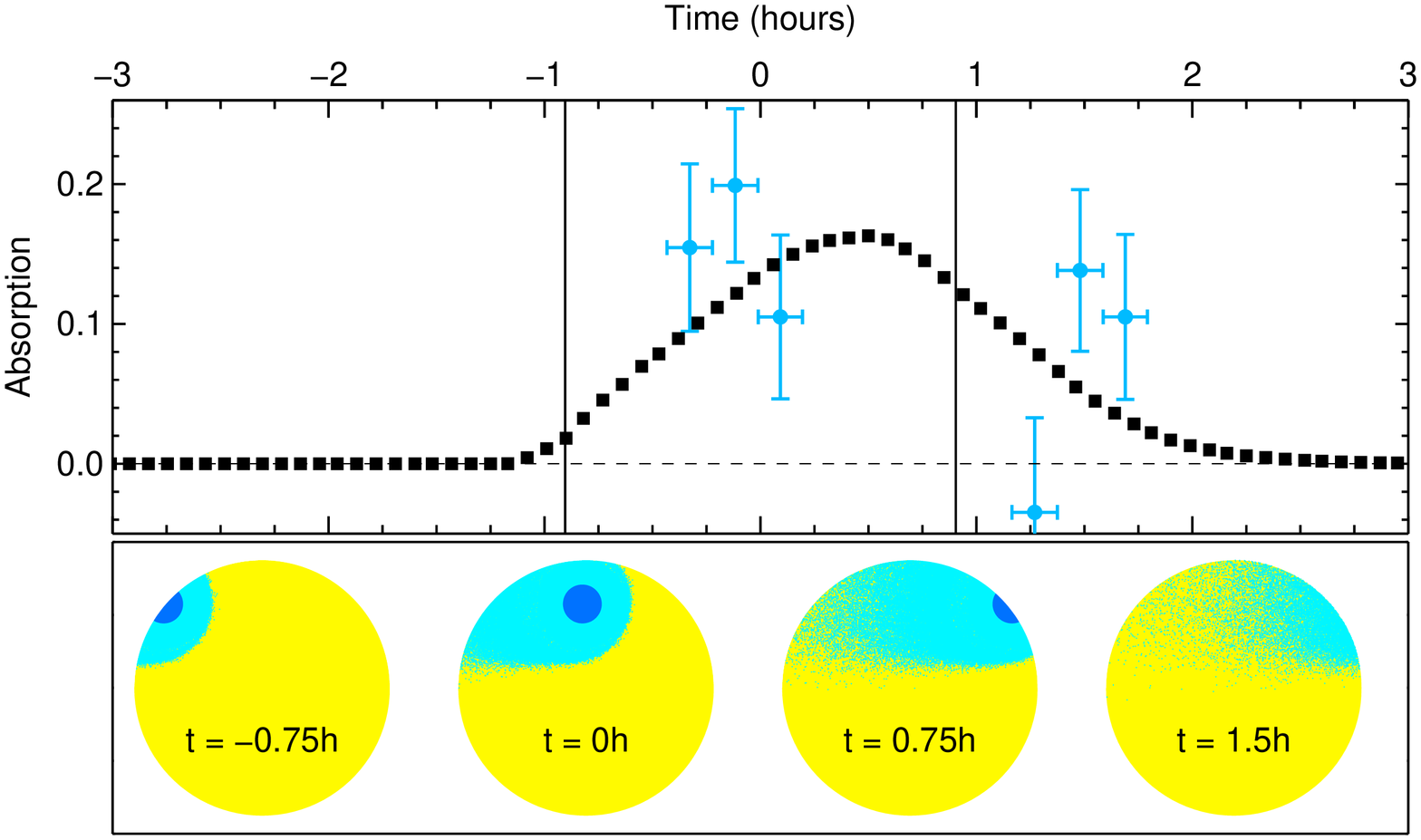}	
\caption[]{Same plot as in Fig.~\ref{sim_tot_abs_HD209} with the absorption calculated over the range -230 to -140\,km\,s$^{-1}$ for HD\,189733b (escape rate $\dot{M}=2\times10^{9}$\,g\,s$^{-1}$; ionizing flux $F_{\mathrm{ion}}=10\,F_\odot$, stellar wind with temperature $T_{\mathrm{wind}}=3\times10^{4}$\,K, bulk velocity $V_{\mathrm{wind}}=200$\,km\,s$^{-1}$, and density $n_{\mathrm{wind}}=2\times10^{4}$cm$^{-3}$). The blue points correspond to the transit and post-transit observations divided into 751~s time exposures (see \citealt{Bourrier2013}). Only particles with observed velocities in the range -230 to -140\,km\,s$^{-1}$ are represented.} 
\label{sim_tot_abs_HD189}
\end{minipage}
\end{figure*}

In this section we study in detail the influence of the cometary-tail structure on the profile of the absorption (Sect.~\ref{com_tail}). In addition to the spatial extension of the cloud, which may reach several tens of planetary radii, the successive regions of the hydrogen cloud crossing in front of the stellar disk during and after the planetary transit vary both in size and optical depth. As a result, the amount of stellar light absorbed by neutral hydrogen atoms in the Lyman-$\alpha$ line may change significantly with time during the transit (\citealt{VM2004}; \citealt{Schneiter2007}; \citealt{Ehrenreich2008}; \citealt{Ekenback2010}; \citealt{Lecav2012}). In Figs.~\ref{sim_tot_abs_HD209} and \ref{sim_tot_abs_HD189} we study the variations of the total absorption in the velocity range of the blue wing signatures defined in Sect.~\ref{observations} (-130 to -40\,km\,s$^{-1}$ for HD\,209458b and -230 to -140\,km\,s$^{-1}$ for HD\,189733b). These variations are quite similar for both planets, even if the absorption profiles are due to populations of neutral hydrogen atoms moving at different velocities. During the first half of the transit, the absorption depth increases steadily because the cloud occults ever more of the planet over time. The highest depth is not reached at mid-transit but at some time around three quarters of the planetary transit duration, when the foremost layers of the hydrogen cloud begin their egress while the farthest regions begin their ingress. The band of stellar surface in the path of the extended cloud is then occulted at its maximum. Soon after the fourth contact, the planetary disk does not occult the star anymore, but the cloud continues to significantly absorb the stellar light for 1.5 to 2 hours. The variations of the absorption within the transit duration show the need to compare measurements with theoretical fluxes averaged over the same time windows as the observations, and not calculated at the center of the transit only.

\subsection{Probing the structure of the hydrogen cloud}

\begin{figure*}
\centering
\begin{minipage}[b]{0.9\textwidth}	
\includegraphics[angle=-90,width=\columnwidth]{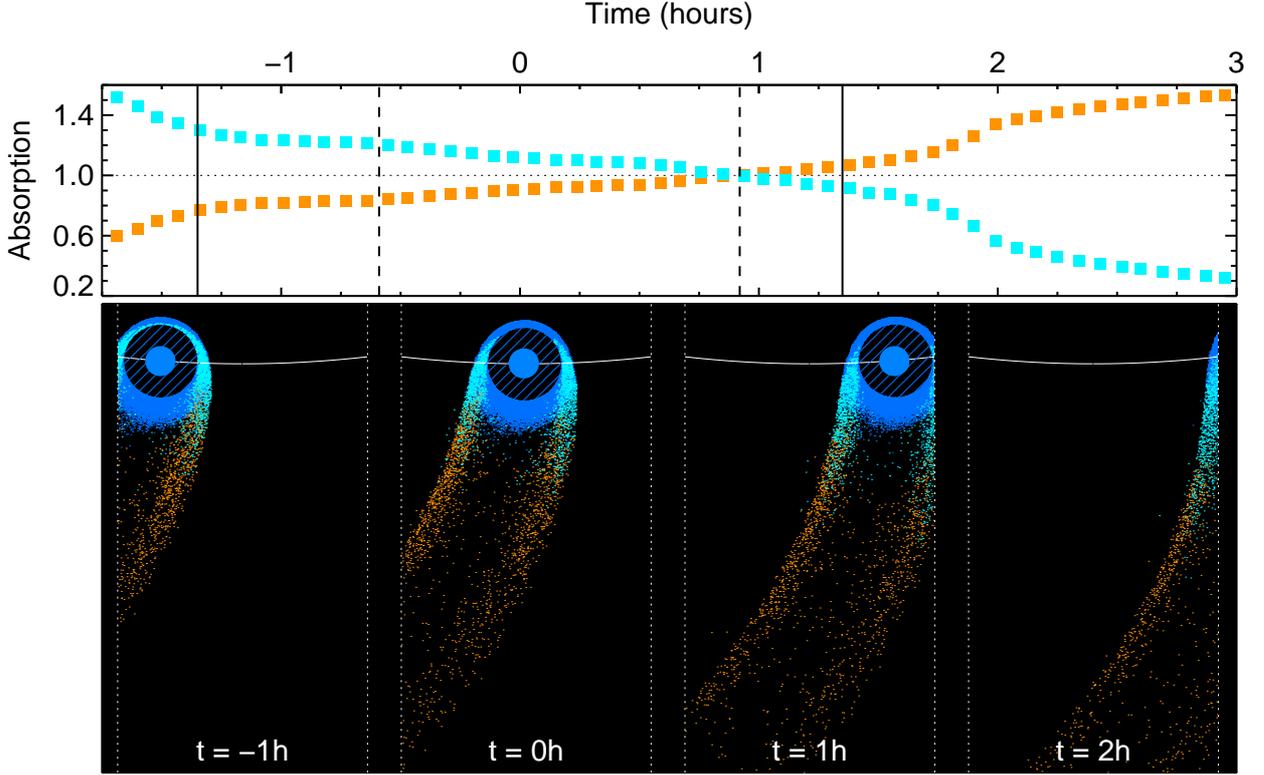}
\end{minipage}
\caption[]{\textit{Top panel:} Absorption over the range -130 to -85\,km\,s$^{-1}$ (orange symbols) and -85 to -40\,km\,s$^{-1}$ (blue symbols), normalized by the absorption over the whole velocity interval, which corresponds to the interval of the absorption signature detected by \citet{VM2003}. The velocities are the projections on the star/Earth line of sight. Each point corresponds to a time step in a simulation of atmospheric escape from HD\,209458b (escape rate $\dot{M}=10^{9}$\,g\,s$^{-1}$; ionizing flux $F_{\mathrm{ion}}=3\,F_\odot$). Vertical solid lines show the beginning and end of ingress and egress of the transit, and vertical dashed lines show the beginning and end of STIS transit observations. \textit{Bottom panel:} Views of the gas in the orbital plane that moves with velocities in the intervals described above (orange and light blue), and below $\pm$40\,km\,s$^{-1}$ (deep blue). The atmosphere below the exobase (striped disk) is uniformely filled with high-density neutral hydrogen gas. The hydrogen atoms are accelerated by radiation pressure and reach higher velocities farther away from the planet (blue disk). At different times during and after the transit, different areas of the cloud cross in front of the stellar disk (the field of view from the Earth is delimited by the dashed white lines). As a result, the relative absorption at low (high) velocities decreases (increases) with time.} 
\label{dual_abs_HD209}
\end{figure*}

\begin{figure*}
\centering
\begin{minipage}[b]{0.9\textwidth}	
\includegraphics[angle=-90,width=\columnwidth]{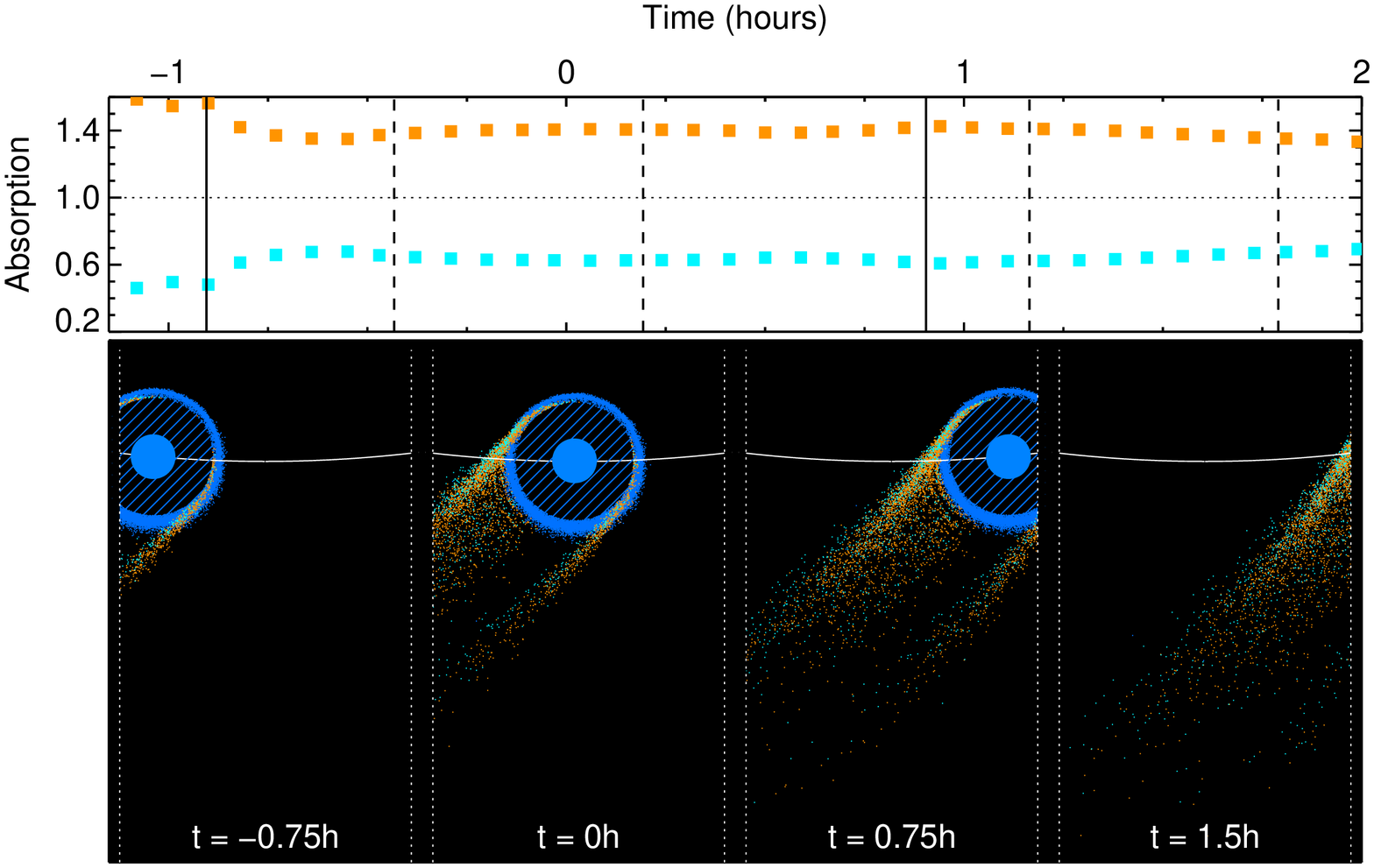}
\end{minipage}
\caption[]{Same plot as in Fig.~\ref{dual_abs_HD209}. \textit{Top panel:} Absorption over the range -230 to -175\,km\,s$^{-1}$ (orange symbols) and -175 to -140\,km\,s$^{-1}$ (blue symbols) for a simulation of atmospheric escape from HD\,189733b (escape rate $\dot{M}=2\times10^{9}$\,g\,s$^{-1}$; ionizing flux $F_{\mathrm{ion}}=3\,F_\odot$; stellar wind with velocity $V_{\mathrm{wind}}=200$\,km\,s$^{-1}$, temperature $T_{\mathrm{wind}}=3\times10^{4}\,K$ and density $n_{\mathrm{wind}}=2.10^{4}$\,cm$^{-3}$). Vertical dashed lines show the beginning and end of STIS transit and post-transit observations. An absorption signature was detected by \citet{Lecav2012} in the blue wing of the Lyman-$\alpha$ line between -230 and -140\,km\,s$^{-1}$. \textit{Bottom panel:} Views of the gas in the orbital plane that moves in the velocity intervals described above (orange and light blue), and between -140\,km\,s$^{-1}$ and 40\,km\,s$^{-1}$ (deep blue). The cometary tail is mostly composed of high-velocity atoms accelerated by interactions with the stellar wind. These atoms keep their Maxwellian velocity distribution as they move farther away from the planet. As a result, the relative absorption over any velocity interval is roughly constant with time. The relative level of the absorption depends on the proportion of atoms in the Maxwellian distribution for a given velocity interval.} 
\label{dual_abs_HD189}
\end{figure*}

It may be possible to probe the structure of the hydrogen cloud by studying the evolution of the velocity structure of the absorption profile as a function of time during and after the planet transit (see \citealt{Bourrier2013}). To distinguish between the 'geometric' variations related to variations of the occulting area and opacity of the cloud in front of the stellar disk (Sect.~\ref{time_var}), and the spectral variations related to the velocity structure of the atoms in the cometary tail, we calculated the total absorptions in two complementary velocity intervals at different times during and after the transit and normalized them by the absorption over the whole interval. This gives an estimate of the proportion of opaque gas absorbing in one or the other interval (with respect to the whole interval) at a given time. For HD\,209458b the different parts of the tail are composed of hydrogen atoms that move in different velocity ranges because the cloud is shaped by radiation pressure. Atoms farther away from the planet have been accelerated over a longer period of time and reach higher velocities (Sect.~\ref{evol_time} and Fig.~\ref{orbital_plane}). Fig.~\ref{dual_abs_HD209} shows that the simulated absorption at low negative velocities is stronger at the beginning of the transit, because most of the atoms that occult the stellar disk are close to the planet. As the transit continues, the regions of the cloud that cross the stellar disk are farther away from the planet and the absorption over higher negative velocities increases accordingly. After the end of the planetary transit, the absorption is mainly due to atoms moving at high velocities in the extremity of the tail. The signal-to-noise ratio in the STIS observations of HD\,209458b does not allow the detection of such variations. Nonetheless, with a better signal-to-noise ratio, these variations could be a way to distinguish between atoms accelerated by radiation pressure and those accelerated by interactions with stellar wind protons. Indeed, the same analysis made for HD\,189733b shows that there are no such variations of relative absorption between low and high-velocity atoms (Fig.~\ref{dual_abs_HD189}). To reproduce the STIS observations of HD\,189733b, the cometary tail must be composed mostly of atoms accelerated by the stellar wind (Sect.~\ref{HD189}). These atoms are accelerated by charge-exchange near the planet and then continue to move with about the same velocity since at high velocities they are less subjected to radiation pressure (Sect.~\ref{rad_press}). Because the absorption profile generated by these high-velocity atoms has the same Maxwellian distribution as the stellar wind and because this distribution is preserved in the different parts of the cloud, there is no variation of the absorption with time in different velocity ranges. Note that this conclusion is not consistent with the marginal detection of absorption variation during the transit as presented by \citet{Bourrier2013}. However, better and more sensitive observations are needed to investigate the reason for this possible discrepancy.\\
The cometary tail extending from HD\,209458b is less bent toward the star than the tail extending from HD\,189733b (see Fig.~\ref{dual_abs_HD209} and Fig.~\ref{dual_abs_HD189}). In the first case, the hydrogen gas escapes the atmosphere with a velocity dominated by the planet tangential velocity and is then shaped by radiation pressure. As a result, the cometary tail of HD\,209458b is shifted toward the planet orbital motion. For HD\,189733b, the velocity distribution of the atoms in the tail, mostly accelerated by interactions with protons, is centered on the radial bulk velocity of the stellar wind. Our model showed that the STIS observations of HD\,189733b do not allow for a significant population of atoms accelerated by radiation pressure, but in another context a dual situation where two cometary tails with different orientations and velocity distributions coexist could be possible.

\section{Conclusion}
\label{conclu}

We presented a numerical 3D particle model developed to simulate the escape of neutral hydrogen from the atmosphere of a planet and calculate the resulting transmission spectrum in Lyman-$\alpha$. The main parameters of this model are the neutral hydrogen escape rate and the ionizing flux from the star. We showed that the dynamics of the escaping atoms is strongly constrained by the velocity-dependent stellar radiation pressure, which can be well estimated using Lyman-$\alpha$ line profiles reconstructed from HST/STIS observations. The escaping gas can be additionally accelerated by interactions with high-velocity protons of the stellar wind, characterized by three more parameters (bulk velocity, temperature, and density of the protons at the orbit of the planet). We took into account the self-shielding of the hydrogen atoms from stellar photons and protons, which may significantly modify the structure of the gas cloud. The theoretical absorption includes thermal and natural broadening and can be calculated for any period of time during or after the planetary transit. It is processed to be compared with STIS observations in the Lyman-$\alpha$ line. \\
We applied our model to the hot-Jupiters HD\,209458b and HD\,189733b to constrain the physical conditions in their exosphere, using the observed atmospheric absorption signatures detected in the blue wing of the Lyman-$\alpha$ line (\citealt{VM2003} and \citealt{Lecav2012}). For HD\,209458b, the observations can be explained by a radiative blow-out that shapes the escaping hydrogen atoms into an extended cometary tail. The observations are well-fitted with two possible scenarios: a hydrogen escape rate $\dot{M}\sim10^{9}$\,g\,s$^{-1}$ and an ionizing flux $F_{\mathrm{ion}}\sim3\,F_\odot$, or escape rates in the range \mbox{$6\times10^{9}$ -- $10^{11}$\,g\,s$^{-1}$} and ionizing fluxes $F_{\mathrm{ion}}\sim4\,F_\odot$ (Fig.~\ref{chi2-HD209}). Shorter lifetimes of the hydrogen atoms compensate for higher escape rates. The higher velocities of the hydrogen atoms in the exosphere of HD\,189733b, as observed in 2011, can be explained by interactions between the planetary and stellar winds, with the bulk velocity and temperature of the protons constrained to $V_{\mathrm{wind}}=200\pm20$\,km\,s$^{-1}$ and $T_{\mathrm{wind}}\sim3\times10^{4}\,K$. We inferred $1\sigma$ error bars on the ionizing flux (6 -- 23$\,F_\odot$), the escape rate ($4\times10^{8}$ -- $10^{11}$\,g\,s$^{-1}$), and the proton density ($10^{3}$ -- $10^{7}$\,cm$^{-3}$), again with higher ionizing fluxes corresponding to higher escape rates or proton density (Fig.~\ref{chi2_HD189_flux10} to \ref{chi2_HD189_flux}). Most notably, proton densities above $\sim3\times10^{5}$\,cm$^{-3}$ lead to an escape-limited \textit{saturation regime} in which most of the hydrogen atoms that escape the atmosphere interact with the stellar wind and the observed absorption is independent on the proton density. For a given measurement of the absorption spectrum outside of this regime, higher escape rates corresp ond to higher ionizing fluxes, or lower densities of the stellar wind. The non-detection of absorption in the Lyman-$\alpha$ observations of HD\,189733 in 2010 can be explained either by escape rates of about an order of magnitude lower than in 2011, or by the absence of any significant stellar wind. \\
Our simulations revealed that in case of a radiative blow-out we could expect to observe time-variations of the velocity profile of the absorption during and after the transit. In contrast, no variations were found in an extended exosphere dominated by interactions with the stellar wind. In addition to being a way for probing the spatial and velocity structure of the escaping gas cloud, this may help in distinguishing the influence of radiation pressure and hydrogen/proton interactions in future observations. The possibility that a cometary tail extends far away from the planet and is responsible for a significant fraction of the absorption, along with spectro-temporal variations of the absorption profile with time, shows the need for 3D modeling of the atmospheric escape.

\begin{acknowledgements}
We thank A. Vidal-Madjar, H. Dupuy and G. F\'eraud for fruitful discussions. We are also grateful to the anonymous referee for his thoughtful comments.\\
This work was based on observations made with the NASA/ESA Hubble Space Telescope, obtained at the Space Telescope Science Institute, which is operated by the Association of Universities for Research in Astronomy, Inc., under NASA contract NAS 5-26555. The authors acknowledge financial support from the Centre National d'\'Etudes Spatiales (CNES). The authors also acknowledge the support of the French Agence Nationale de la Recherche (ANR), under program ANR-12-BS05-0012 "Exo-Atmos". This work has also been supported by an award from the Fondation Simone et Cino Del Duca. 
\end{acknowledgements}

\bibliographystyle{aa} % style aa.bst
\bibliography{biblio} % your references Yourfile.bib

\end{document}